\documentclass[prd, twocolumn, showpacs,aps,nofootinbib,floatfix,amsmath,amssymb]{revtex4}
\usepackage{graphicx}
\usepackage{soul}
\usepackage{color}
\usepackage[mathscr]{euscript}
\usepackage[usenames,dvipsnames]{xcolor}
\usepackage{amsmath}
\allowdisplaybreaks
\begin{document}

\makeatletter
\newbox\slashbox \setbox\slashbox=\hbox{$/$}
\newbox\Slashbox \setbox\Slashbox=\hbox{\large$/$}
\def\pFMslash#1{\setbox\@tempboxa=\hbox{$#1$}
  \@tempdima=0.5\wd\slashbox \advance\@tempdima 0.5\wd\@tempboxa
  \copy\slashbox \kern-\@tempdima \box\@tempboxa}
\def\pFMSlash#1{\setbox\@tempboxa=\hbox{$#1$}
  \@tempdima=0.5\wd\Slashbox \advance\@tempdima 0.5\wd\@tempboxa
  \copy\Slashbox \kern-\@tempdima \box\@tempboxa}
\def\FMslash{\protect\pFMslash}
\def\FMSlash{\protect\pFMSlash}
\def\miss#1{\ifmmode{/\mkern-11mu #1}\else{${/\mkern-11mu #1}$}\fi}

\newcommand{\psum}[1]{{\sum_{ #1}\!\!\!}'\,}
\makeatother

\title{Quark-flavor-changing Higgs decays from a universal extra dimension}

\author{Carlos M. Farrera, Alejandro Granados-Gonz\'alez, H\'ector Novales-S\'anchez, and J. Jes\'us Toscano}
\affiliation{Facultad de Ciencias F\'isico Matem\'aticas, Benem\'erita Universidad Aut\'onoma de Puebla, Apartado Postal 1152 Puebla, Puebla, M\'exico.}

\begin{abstract}
Kaluza-Klein fields characterizing, from a four-dimensional viewpoint, the presence of compact universal extra dimensions would alter low-energy observables through effects determined by some compactification scale, $R^{-1}$, since the one-loop level, thus being particularly relevant for physical phenomena forbidden at tree level by the Standard Model. The present paper explores, for the case of one universal extra dimension, such new-physics contributions to Higgs decays $h^{(0)}\to q^{(0)}_\alpha q^{(0)}_\beta$, into pairs of quarks with different flavors, a sort of decay process which, in the Standard Model, strictly occurs at the loop level. Finite results, decoupling as $R^{-1}\to \infty$, are calculated. Approximate short expressions, valid for large compactification scales, are provided. We estimate that Kaluza-Klein contributions lie below predictions from the Standard Model, being about 2 to 3 orders of magnitude smaller for compactification scales within $1.4\,{\rm TeV}<R^{-1}<10\,{\rm TeV}$.
\end{abstract}


\maketitle


\section{Introduction}
\label{int}
The Standard Model~\cite{Glashow,Weinberg,Salam}, a field theory governed by the Poincar\'e group ${\rm ISO}(1,3)$ and by the gauge-symmetry group ${\rm SU}(3)_C\times{\rm SU}(2)_L\times{\rm U}(1)_Y$, is the current best fundamental description of nature~\cite{PDG}. In particular, the measurement, in 2012, of a scalar particle with mass $\sim125\,{\rm GeV}$, at the {\it Large Hadron Collider} (LHC) by the ATLAS and CMS Collaborations~\cite{ATLASHiggs,CMSHiggs}, as well as further studies on its properties~\cite{ATLASHb2013,CMSHb2013,EllYou,DjMo,ATLASHb2015,ATLASandCMSHb2015,ATLASHb2020}, display good agreement with the Standard-Model Higgs boson and thus seem to confirm the Standard-Model scalar sector to be the correct formulation behind the origin of mass in the known universe. Nonetheless, this minimal scalar sector is not, by any means, the only option, since candidates from several Standard-Model extensions, involving richer scalar sectors, exist. In this context, investigations aimed at the precise characterization and quantification of Higgs-boson interactions, in the presence of new physics beyond the Standard Model, are interesting and relevant. \\

Among the variety of Standard-Model extensions, the present investigation is carried out within the framework of a field theory defined on an extra-dimensional spacetime. Initially pointing towards unification of fundamental interactions~\cite{Nordstrom,Kaluza,Klein} and then developed in parallel with string theories~\cite{Veneziano,Nielsen,KoNi1,KoNi2,Nambu,Susskind1,Susskind2,Lovelace,Ramond,WeZu,SchSch,Schwarz,GS,GS1,GHMR1,GHMR2,GHMR3,Yau,CHSW,Witten,Polchinski,HoWi1,HoWi2,Maldacena}, extra-dimensional formulations gained great phenomenological appeal when the plausibility of {\it large}, ${\rm TeV}$-sized, {\it extra dimensions} was pointed out in Refs.~\cite{A,ADD,AADD}. Geometrical motivations also inspired models of {\it warped extra dimensions}, aimed at the hierarchy problem~\cite{RS1,RS2}. In the present work, another well-known extra-dimensional framework is considered, commonly referred to as {\it universal extra dimensions}~\cite{ACD1,AppDo,RizzoUED,ADPY,CMS,SeTa1,SeTa2,AppYee,BSW,BPSW,HoPr,NT,CGNT,Servant} and characterized by the assumption that all the dynamic variables defining a given field formulation depend on the coordinates of the whole spacetime with spatial extra dimensions. The Standard Model in $4+n$ universal extra dimensions is a field theory defined in terms of extra-dimensional dynamic fields which are replicas of the fields constituting the four-dimensional Standard Model (4DSM). Moreover, spacetime symmetry ${\rm ISO}(1,3+n)$ is assumed to hold at very high energies, whereas, within the same energy range, the theory is assumed to be invariant with respect to the gauge-symmetry group\footnote{Here, ${\cal M}^{4+n}$ means that the gauge group is defined on a Minkowski-like spacetime with $4+n$ dimensions.} ${\rm SU}(3,{\cal M}^{4+n})_C\times{\rm SU}(2,{\cal M}^{4+n})_L\times{\rm U}(1,{\cal M}^{4+n})_Y$, which is the same as that of the 4DSM in the sense of its gauge generators, but differs in that it is defined rather on the $(4+n)$-dimensional spacetime. \\

While the dynamic variables and symmetries of these 4DSM extensions are defined on spacetimes with extra dimensions, the experimentally-supported~\cite{DFK} assumption that extra dimensions are compact allows explicit integration of the extra-dimensional coordinates in the action, thus yielding effective field theories, known as {\it Kaluza-Klein} (KK) {\it theories}, whose dynamic variables, dubbed {\it KK modes}, depend only on 4 spacetime dimensions~\cite{DDG,MPR,RizzoED}. For each extra-dimensional field of the theory, here generically denoted by $\chi$, an infinite set of KK modes $\chi^{(\underline{k})}$ is generated, each labeled by a discrete multi-index $(\underline{k})=(k_1,k_2,\ldots,k_n)$. Among all the KK fields, a subset of fields known as {\it KK zero modes}, $\chi^{(\underline{0})}$, are identified as the dynamic variables of the 4DSM, whereas the rest of the fields, called {\it KK excited modes}, incarnate new dynamic variables characterizing, from a four-dimensional viewpoint, the presence of extra dimensions. An appealing feature of KK theories originated in field models of universal extra dimensions is the small number of added parameters, which are a high-energy {\it compactification scale}, $R^{-1}$, and the number of extra dimensions, $n$. Moreover, these models offer {\it dark-matter candidates}~\cite{SeTa1,SeTa2,BKP,FKKMP}, which would be the first KK excited mode of either the photon or the neutrino. \\

In passing from $4+n$ dimensions to four, the {\it KK mass-generating mechanism} operates~\cite{LMNT2}, giving each KK excited mode a mass, which then receives a further contribution through the {\it Englert-Higgs mechanism}~\cite{EnBr,PWHiggs1,PWHiggs2} (EHM) of the 4DSM, provoked by the assumption that the zero-mode Higgs doublet triggers, in complicity with the zero-mode scalar potential, {\it spontaneous symmetry breaking}. As usual, the Higgs field $h^{(0)}$ emerges as a byproduct. 
Detailed discussions on the role of the EHM in the Standard Model in universal extra dimensions and its associated KK theory can be found in Refs.~\cite{CGNT,MNT}. Phenomenological investigations on Higgs physics in this extra-dimensional context have been carried out as well~\cite{AppYee,Petriello,NoTo1,BBBKP,DKandSR,KDandJ}. An important feature of field models with universal extra dimensions is that low-energy Green's functions, and thus 4DSM observables, receive their very first corrections at the loop level, which is a consequence of conservation of extra-dimensional momentum~\cite{ACD1}. Therefore, physical processes and observables forbidden in the 4DSM at the tree level are particularly important for this kind of new physics. With this in mind, the present paper explores the effects produced by the KK modes originating in the five-dimensional Standard Model (5DSM) on the decays $h^{(0)}\to q^{(0)}_\alpha q^{(0)}_\beta$, of the Higgs boson into 4DSM quark pairs, with quark-flavor change, that is with $\alpha\ne\beta$, and where final-state quarks can be either $u$- or $d$-type. The geometry of the compact extra dimension is assumed to be an orbifold $S^1/Z_2$, with radius $R$. We find exact results that are ultraviolet (UV) finite, written in terms of Passarino-Veltman scalar functions~\cite{PassVe,HooVe} and which decouple as $R^{-1}\to\infty$. By assuming the extra dimension to be tiny, we find approximate expressions, defined by elementary functions of masses. These analytical expressions are then implemented to the specific decay processes $h^{(0)}\to u^{(0)}c^{(0)}$, $h^{(0)}\to d^{(0)}s^{(0)}$, $h^{(0)}\to s^{(0)}b^{(0)}$, and $h^{(0)}\to b^{(0)}d^{(0)}$. We conclude that the effects from this extra-dimensional new physics lie about 2 to 3 orders of magnitude below 4DSM contributions, for a compatification scale in the interval $1.4\,{\rm TeV}<R^{-1}<10\,{\rm TeV}$. \\

The paper has been organized as follows: in Section~\ref{theory} a brief discussion on the theoretical framework is carried out, including explicit expressions of Lagrangian terms which are relevant for the main calculation; the analytical calculation of the contributions from the whole KK theory to the branching ratio for $h^{(0)}\to q^{(0)}_\alpha q^{(0)}_\beta$ is presented and discussed in Section~\ref{analytical}, where approximate expressions, valid for very large compactification scales, are derived; these results are implemented to specific decays in Section~\ref{numbers}, where numerical estimations and a discussion are provided; finally, our conclusions are presented in Section~\ref{conclusions}.


\section{The five-dimensional Standard Model and its Kaluza-Klein theory}
\label{theory}
In this section, a brief description of the model is presented. We refer the reader interested in the details of this formulation to Refs.~\cite{NT,CGNT,LMNT2,MNT,LMNT1}, in which thorough discussions on the matter can be found. A main purpose of the discussion at hand is the presentation of KK Lagrangian terms from which Feynman rules follow in order to write down the one-loop Feynman diagrams contributing to the amplitude for the decay process $h^{(0)}\to q^{(0)}_\alpha q^{(0)}_\beta$. \\

The 5DSM is a field formulation set on a five-dimensional Minkowski-like spacetime, characterized by the metric $g^{MN}={\rm diag}(1,-1,-1,-1,-1)$, with capital-letter indices corresponding to the five spacetime coordinates: $M,N=0,1,2,3,5$. The symmetry groups of this theory are assumed to be ${\rm ISO}(1,4)$, for spacetime, and ${\rm SU}(3,{\cal M}^5)_C\times{\rm SU}(2,{\cal M}^5)_L\times{\rm U}(1,{\cal M}^5)_Y$ for gauge symmetry in five spacetime dimensions. We assume that such symmetries govern the physical description at very high energies, while at some lower-energy scale, which we refer to as the {\it compactification scale}, the extra dimension manifests its compact nature, here assumed to have the geometry of the orbifold $S^1/Z_2$, with radius $R$. Under such circumstances, $R^{-1}$ is used to characterize the compactification scale, which we assume to be larger than the electroweak scale $v=246\,{\rm GeV}$. This hierarchy of energy scales is consistent with lower bounds on the compactification scale $R^{-1}$, among which the most stringent limit, derived from LHC data, is $R^{-1}>1.4\,{\rm TeV}$~\cite{DFK}.\\

The 5DSM Lagrangian, ${\cal L}^{\rm SM}_5$, is written as the sum ${\cal L}_5^{\rm SM}={\cal L}^{\rm YM}_5+{\cal L}^{\rm S}_5+{\cal L}^{\rm Y}_5+{\cal L}^{\rm C}_5$, of the {\it Yang-Mills sector} ${\cal L}^{\rm YM}_5$, the {\it scalar sector} ${\cal L}^{\rm S}_5$, the {\it Yukawa sector} ${\cal L}^{\rm Y}_5$, and the {\it currents sector} ${\cal L}^{\rm C}_5$, all of them defined in five spacetime dimensions. The field content comprising the 5DSM is analogue to the set of dynamic variables of the 4DSM, but with all the fields defined in the five spacetime dimensions. The gauge-symmetry group ${\rm SU}(2,{\cal M}^5)_L$ comes along with 3 connections $W^{j}_M(x,\bar{x})$, where $j=1,2,3$ is the gauge index, whereas the Abelian group ${\rm U}(1,{\cal M}^5)_Y$ introduces the gauge field $B_M(x,\bar{x})$. An ${\rm SU}(2,{\cal M}^5)_L$ scalar doublet $\Phi(x,\bar{x})$, with hypercharge $Y_\Phi$, defines the scalar sector. In view of the lack of a proper chiral matrix in five-dimensional spinor-field theories, nonchiral ${\rm SU}(2,{\cal M}^5)_L$ doublets,
\begin{equation}
L_\alpha
=
\left(
\begin{array}{c}
\tilde{\nu}_\alpha(x,\bar{x})
\\
\tilde{l}_\alpha(x,\bar{x})
\end{array}
\right),
\hspace{0.5cm}
Q_\beta
=
\left(
\begin{array}{c}
\tilde{u}_\beta(x,\bar{x})
\\
\tilde{d}_\beta(x,\bar{x})
\end{array}
\right),
\end{equation}
where $\alpha=e,\mu,\tau$ and $\beta=u,c,t$, as well as ${\rm SU}(2,{\cal M}^5)_L$ nonchiral singlets
\begin{equation}
\nu_\alpha(x,\bar{x}),\hspace{0.5cm}l_\alpha(x,\bar{x}),\hspace{0.5cm}u_\beta(x,\bar{x})\hspace{0.5cm}d_\beta(x,\bar{x}),
\end{equation}
where again $\alpha=e,\mu,\tau$ and $\beta=u,c,t$, are introduced. The orbifold geometry of the compact dimension is such that periodicity and defined-parity properties with respect to the extra dimension can be given to the fields, which are then Fourier expanded as
\begin{eqnarray}
\chi(x,\bar{x})=\frac{1}{\sqrt{2\pi R}}\chi^{(0)}(x)+\sum_{k=1}^\infty\frac{1}{\sqrt{\pi R}}\chi_{\rm E}^{(k)}(x)\cos\Big( \frac{k\bar{x}}{R} \Big)
\nonumber \\ 
+\sum_{k=1}^\infty\frac{1}{\sqrt{\pi R}}\chi_{\rm O}^{(k)}(x)\sin\Big( \frac{k\bar{x}}{R} \Big),
\label{KKexp}
\end{eqnarray}
with $x$ and $\bar{x}$ denoting the ordinary four-dimensional coordinates and the coordinate of the extra dimension, respectively.
In this equation, KK modes $\chi^{(0)}$, $\chi_{\rm E}^{(k)}$, $\chi_{\rm O}^{(k)}$ may or may not be present, which is determined by the specific $\bar{x}$-parity property of the five-dimensional field $\chi$ and by the type of field as well. Five-dimensional tensor fields involve either even KK modes $\chi^{(0)}$ and $\chi_{\rm E}^{(k)}$, if $\chi(x,-\bar{x})=+\chi(x,\bar{x})$, or odd KK modes $\chi_{\rm O}^{(k)}$, in case that $\chi(x,-\bar{x})=-\chi(x,\bar{x})$. Moreover, five-dimensional spinor fields include both even and odd KK modes, which are four-dimensional chiral spinors, with chiralities determined by parity of $\chi$~\cite{CDH,PapaSan}.\\

To map the five-dimensional fields into their KK modes, which depend only on four spacetime dimensions, two canonical transformations are implemented~\cite{LMNT1}, one of which characterizes the explicit breaking of extra-dimensional Lorentz symmetry by compactification whereas the other is the aforementioned set of Fourier expansions. The KK modes are the dynamic variables of the theory after compactification. While the spacetime-symmetry group of the KK theory is ${\rm ISO}(1,3)$, gauge symmetry in five dimensions gets split, at four dimensions, into two disjoint sets of transformations which leave the theory invariant, namely, the {\it standard gauge transformations} and the {\it nonstandard gauge transformations}~\cite{NT,LMNT1}. The standard gauge transformations constitute the gauge group ${\rm SU}(3,{\cal M}^4)_C\times{\rm SU}(2,{\cal M}^4)_L\times{\rm U}(1,{\cal M}^4)_Y$, defined in four dimensions of spacetime and under which the KK zero modes of five-dimensional gauge fields are the only ones transforming as four-dimensional gauge fields, thus being recognized as the connections of the KK gauge theory. Nonstandard gauge transformations, on the other hand, do not form a group; they are a manifestation that extra-dimensional gauge symmetry is still present, though not explicitly.\\

After incorporation of the KK modes in the theory, through Eq.~(\ref{KKexp}), the fifth spacetime coordinate, corresponding to the compact dimension, stops labeling degrees of freedom, a role taken over by KK indices $(k)$, and its presence in the Lagrangian gets relegated to trigonometric functions, which can be straightforwardly integrated out from the Action as $S^{\rm SM}_5=\int d^5x\,{\cal L}^{\rm SM}_5(x,\bar{x})=\int d^4x\,{\cal L}^{\rm SM}_{\rm KK}(x)$, where ${\cal L}^{\rm SM}_{\rm KK}(x)=\int_0^{2\pi R} d\bar{x}\,{\cal L}^{\rm SM}_5(x,\bar{x})$ is the four-dimensional effective KK Lagrangian. Even though compactification breaks five-dimensional Lorentz invariance, the KK theory is still invariant under the four-dimensional spacetime symmetry group ${\rm ISO}(1,3)$. In this context, any five-dimensional vector splits into a KK four-vector of ${\rm SO}(1,3)$ and a KK Lorentz scalar. Nonchiral spinors in five dimensions separate into two sets of right-handed and two sets of left-handed four-dimensional spinors, with left-handed spinors constituting ${\rm SU}(2,{\cal M}^4)_L$ doublets and right-handed spinors being singlets under such a gauge group. Furthermore, the KK mechanism generates a nonzero mass $m_{(k)}=k/R$, determined by the compactification scale $R^{-1}$ and the KK index $(k)$, for KK excited modes while leaving zero modes massless. In particular, gauge KK zero modes, protected by gauge symmetry in four dimensions, consistently remain massless, though the associated vector KK excited modes become massive. In the case of KK scalars generated from five-dimensional gauge modes, they remain massless as well, but, contrastingly to the case of zero-mode gauge fields, they are unphysical, sort of like pseudo-Goldstone bosons, in the sense that a gauge choice, with respect to nonstandard gauge transformations, which eliminates them from the KK theory exists~\cite{NT,CGNT,LMNT1}. \\

Once the process described in the previous paragraph has occurred, the resulting KK Lagrangian ${\cal L}^{\rm SM}_{\rm KK}$ comprises the 4DSM and a large set of couplings involving KK zero and excited modes, with the zero modes playing the role of 4DSM fields. To this respect, let us emphatically point out that a judicious endowment of $\bar{x}$-parity properties of five-dimensional fields is crucial to get the 4DSM as the low-energy theory. Even so, note that alternative parity properties have been taken advantage of in models of {\it gauge-Higgs unification}~\cite{HosoConf}. In particular, the 4DSM includes a scalar sector, given in terms of the zero mode ${\rm SU}(2,{\cal M}^4)$ doublet $\Phi^{(0)}$. This zero-mode sector involves the gauge-invariant scalar potential $V(\Phi^{(0)},\Phi^{(0)\dag})=-\mu^2\Phi^{(0)\dag}\Phi^{(0)}+\lambda(\Phi^{(0)\dag}\Phi^{(0)})^2$, with $\mu^2>0$ a quantity with units $({\rm mass})^2$ and $\lambda$ a dimensionless and positive coupling constant. Having a scalar potential with a gauge invariant set of minima, the theory is, as usual, subjected to spontaneous symmetry breaking. The particular minimum determined by the condition $\phi_0^{(0){\rm T}}=(0,v/\sqrt{2})$ is chosen, so that the EHM takes place. Most of KK zero-mode fields $\chi^{(0)}$ become massive with mass $m_{\chi^{(0)}}$, in the standard manner, whereas each KK excited mode receives a mass-term contribution, resulting in a total mass $m_{\chi^{(k)}}$, given by $m^2_{\chi^{(k)}}=m^2_{\chi^{(0)}}+m^2_{(k)}$.\\

For practical reasons, we find it convenient to describe the resulting field content of the KK effective theory. Regarding the set of KK zero modes, we have the 4DSM spectrum which emerges after spontaneous symmetry breaking. There is a massless gauge field $A^{(0)}_\mu$, to be identified as the electromagnetic field, whereas charged vector fields $W^{(0)+}_\mu$ and $W^{(0)-}_\mu$, with mass $m_{W^{(0)}}=gv/2$, and a neutral vector field $Z^{(0)}_\mu$, with mass $m_{Z^{(0)}}=m_{W^{(0)}}/c_W$, appear. Here, $g$ is the dimensionless ${\rm SU}(2,{\cal M}^4)$ coupling constant and $c_W=\cos\theta_W$, with $\theta_W$ the {\it weak mixing angle}. The standard Higgs field $h^{(0)}$, with mass $m_{h^{(0)}}=\sqrt{2\mu^2}$, also emerges, together with the pseudo-Goldstone bosons $G^{(0)+}_W$, $G^{(0)-}_W$, $G_Z^{(0)}$. In the zero-mode lepton sector, there are 3 charged leptons $l^{(0)}_e$, $l^{(0)}_\mu$, $l^{(0)}_\tau$, all of them with electric charge $-e$ and with masses $m_{l^{(0)}_e}$, $m_{l^{(0)}_\mu}$, $m_{l^{(0)}_\tau}$, respectively. Each zero-mode lepton field $l^{(0)}_\alpha$ has an associated zero-mode flavor-neutrino field $\nu^{(0)}_\alpha$. If neutrino singlets are introduced at the five-dimensional level, as we did, zero-mode-neutrino Dirac-mass terms and mixings arise from the EHM. The mass-eigenspinor neutrino fields $\nu^{(0)}_1$, $\nu^{(0)}_2$, $\nu^{(0)}_3$ then respectively have masses $m_{\nu^{(0)}_1}$, $m_{\nu^{(0)}_2}$, $m_{\nu^{(0)}_3}$, in the same manner as it happens in the {\it minimally extended standard model}~\cite{GiKi}. The KK Lagrangian contains six zero-mode quark fields as well, of which three correspond to $u$-type quarks and three are associated to $d$-type quarks. Zero-mode $u$-type quark fields $u^{(0)}_u$, $u^{(0)}_c$, $u^{(0)}_t$, with masses $m_{u^{(0)}_u}$, $m_{u^{(0)}_c}$, $m_{u^{(0)}_t}$, respectively, and all having the same electric charge $2e/3$, are defined by standard changes of bases, which are also utilized to define the $d$-type quark fields $d^{(0)}_d$, $d^{(0)}_s$, $d^{(0)}_b$, with masses $m_{d^{(0)}_d}$, $m_{d^{(0)}_s}$, $m_{d^{(0)}_b}$ and with electric charge $-e/3$. All zero-mode-fermion masses proceed from the zero-mode Yukawa sector, after transforming spinor fields into the mass-eigenspinor basis by means of the standard biunitary transformation~\cite{CheLi}.\\

The set of KK excited modes is larger than that of zero modes, which was just described, and, furthermore, the procedure to get to the mass-eigenfields basis is more intricate. The specifics of such a procedure can be found in Refs.~\cite{CGNT,MNT}. To describe the KK-excited-mode-field content of the KK theory, we consider some fixed KK index $(k)\ne(0)$. For any such KK index, there is a field spectrum, which replicates for all other KK indices.
Recall that the mass of a KK excited mode $\chi^{(k)}$, associated to the zero-mode field $\chi^{(0)}$, is given by $m^2_{\chi^{(k)}}=m^2_{\chi^{(0)}}+m^2_{(k)}$, where $m_{\chi^{(0)}}$ is generated by the EHM and $m_{(k)}$ comes from the KK mass-generating mechanism. The theory involves a massive vector field $A^{(k)}_\mu$, which is a KK excited mode corresponding to the photon field. The KK excited-mode field spectrum also includes KK charged vector fields $W_\mu^{(k)+}$ and $W_\mu^{(k)-}$, linked to the 4DMS $W$ bosons. Furthermore, a $Z$-boson KK-excited vector field $Z^{(k)}_\mu$, which is electrically neutral, is defined. A KK excited mode $h^{(k)}$, for the Higgs boson, also emerges. After implementation of diagonalization procedures, the pseudo-Goldstone KK excited modes $G^{(k)+}_W$, $G^{(k)-}_W$, $G^{(k)}_Z$ are defined, together with physical charged scalars $W^{(k)+}$, $W^{(k)-}$, with mass $m_{W^{(k)}}$, and a neutral scalar field $Z^{(k)}$, with mass $m_{Z^{(k)}}$. The inclusion of two different sets of nonchiral five-dimensional spinors, in different representations of ${\rm SU}(2,{\cal M}^5)_L$, for each fermion family produces, at the level of four dimensions, two mass-degenrate KK-excited-mode nonchiral spinors. For instance, there are two KK excited muon fields, namely, $l^{(k)}_{1,\mu}$ and $l^{(k)}_{2,\mu}$, which are nonchiral four-dimensional spinors. And the same goes for all other KK-excited-mode lepton and quark fields. Thus, we have six flavor lepton KK fields $l^{(k)}_{1,\alpha}$ and $l^{(k)}_{2,\alpha}$, since $\alpha=e,\mu,\tau$, six neutrino KK fields $\nu^{(k)}_{1,j}$ and $\nu^{(k)}_{2,j}$, for $j=1,2,3$, six $u$-type KK quarks $u^{(k)}_{1,\beta}$ and $u^{(k)}_{2,\beta}$, with $\beta=u,c,t$, and six $d$-type quarks $d^{(k)}_{1,\gamma}$ and $d^{(k)}_{2,\gamma}$, because $\gamma=d,s,b$. \\

While gauge symmetry is essential for the construction of realistic fundamental descriptions of nature, gauge fixing is an imperious step for the quantization of any gauge field theory~\cite{PeSch,HenTe}. The phenomenological calculation to be performed is characterized by Feynman diagrams, each separately generating a gauge-dependent contribution. Since $S$-matrix elements are required to be independent of the gauge choice, such a gauge dependence must vanish when summing all contributing diagrams together. With this in mind, we address gauge fixing of the KK theory. The presence, at the four-dimensional level, of the standard and the nonstandard gauge transformations opens the possibility of implementing two independent gauge-fixing procedures, one per each set of transformations, thus eliminating the whole symmetry from the KK theory. To remove invariance under the standard gauge transformations we choose the {\it unitary gauge}, so zero-mode pseudo-Goldstone bosons shall be absent. In the case of the nonstandard gauge transformations, on the other hand, we consider the so-called {\it nonlinear gauges}~\cite{MRS,MoRe,RoBa,MeTo}. The Becchi-Rouet-Stora-Tyutin quantization~\cite{BRS1,BRS2,Tyutin}, taking place within the {\it field-antifield formalism}~\cite{GPS,BaVi1,BaVi2,BaVi3,BaVi4,BaVi5}, provides a suitable context to quantize gauge systems by a procedure that, from an operative perspective, reduces to the specification of general {\it gauge-fixing functions}. In such a context, the standard linear-gauge-fixing approach can be implemented~\cite{FLS}. Nonetheless, for the present calculation we opted for a nonlinear set of gauge-fixing functions, which are given by
\begin{eqnarray}
f^{(k)j}&=&{\cal D}^{(0)jm}_\mu W^{(k)m\mu}-\xi m_{(k)}W^{(k)j}_{\rm G}
\nonumber \\ &&
+ig\xi\Big( \Phi^{(k)\dag}\frac{\sigma^j}{2}\Phi^{(0)}-\Phi^{(0)\dag}\frac{\sigma^j}{2}\Phi^{(k)} \Big),
\end{eqnarray}
\begin{eqnarray}
f^{(k)}&=&\partial_\mu B^{(k)\mu}-\xi m_{(k)}B^{(k)}_{\rm G}
\nonumber \\ &&
+\frac{ig'Y_\phi}{2}\xi\big( \Phi^{(k)\dag}\Phi^{(0)}-\Phi^{(0)\dag}\Phi^{(k)} \big).
\end{eqnarray}
These equations define the {\it gauge-fixing sector}
\begin{equation}
{\cal L}^{{\rm GF}(k)}_{\rm SM}=-\frac{1}{2\xi}\sum_{k=1}^\infty\big( f^{(k)j}f^{(k)j}+f^{(k)}f^{(k)} \big),
\end{equation}
characterized by the gauge-fixing parameter, $\xi$. Nonlinear gauge-fixing schemes like this have been utilized before in extra-dimensional contexts~\cite{NT,CGNT,MNT,NT1,GNT}, while their use has been of profit in the SM~\cite{Boudjema,HPTT} and SM extensions as well~\cite{HPTT1,MTTR,RTT,HoTo,HHPT}. Within this gauge-fixing approach, we carry out the calculation in the Feynman-'t Hooft gauge, which corresponds to the value $\xi=1$.\\

The full KK effective Lagrangian, found after implementation of compactification and the EHM, includes the whole 4DSM, but also contains a plethora of couplings in which KK excited modes take part. The Lagrangian terms and/or Feynman rules of the 4DSM are available in the literature~\cite{CheLi,Langacker}, so we rather focus on Lagrangian terms in which KK excited-mode fields participate. From the sum ${\cal L}_{\rm KK}^{\rm S}+{\cal L}_{\rm KK}^{{\rm GF}(k)}$, which combines the KK scalar sector and the KK-excited-mode gauge-fixing sector, we have the Lagrangian terms
\begin{equation}
{\cal L}_{h^{(0)}W^{(k)}W^{(k)}}=g\,m_{W^{(0)}}h^{(0)}W^{(k)-}_\mu W^{(k)+\mu},
\end{equation}
\begin{eqnarray}
&&
{\cal L}_{h^{(0)}G^{(k)}_WG^{(k)}_W}=-\frac{g\,m_{W^{(0)}}}{2m^2_{W^{(k)}}}
\nonumber \\ && 
\hspace{1.5cm}
\times
(m^2_{h^{(0)}}+2m^2_{W^{(k)}})
h^{(0)}G^{(k)+}_WG^{(k)-}_W,
\end{eqnarray}
\begin{eqnarray}
&&
{\cal L}_{h^{(0)}W^{(k)}_{\rm s}W^{(k)}_{\rm s}}=-\frac{g}{2m_{W^{(0)}}m^2_{W^{(k)}}}(2m^2_{W^{(0)}}m^2_{W^{(k)}}
\nonumber \\ &&
\hspace{1.7cm}
+m^2_{(k)}m^2_{h^{(0)}})h^{(0)}W^{(k)+}W^{(k)-},
\end{eqnarray}
\begin{equation}
{\cal L}_{h^{(0)}W^{(k)}G^{(k)}_W}=-\frac{g\,m_{W^{(0)}}}{m_{W^{(k)}}}W^{(k)-}_\mu G^{(k)+}_W\partial^\mu h^{(0)}+{\rm H.c.},
\end{equation}
\begin{equation}
{\cal L}_{h^{(0)}W^{(k)}W^{(k)}_{\rm s}}=\frac{g\,m_{(k)}}{m_{W^{(k)}}}W^{(k)-}_\mu W^{(k)+}\partial^\mu h^{(0)}+{\rm H.c.},
\end{equation}
\begin{equation}
{\cal L}_{h^{(0)}G_W^{(k)}W_{\rm s}^{(k)}}=\frac{g\,m_{(k)}m^2_{h^{(0)}}}{2m^2_{W^{(k)}}}G^{(k)+}_WW^{(k)-}h^{(0)}+{\rm H.c.},
\end{equation}
which comprise all the interactions among the Higgs boson and KK excited modes of charged boson fields. \\

The sum ${\cal L}_{\rm KK}^{\rm Y}+{\cal L}_{\rm KK}^{\rm C}$, of the KK Yukawa and Currents sectors, defines the Lagrangian terms
\begin{eqnarray}
&&
{\cal L}_{h^{(0)}u^{(k)}_\alpha u^{(k)}_\alpha}=
\nonumber \\ && \hspace{0.5cm} 
-g\sin(2\theta_{u^{(k)}_\alpha})\frac{m_{u^{(0)}_\alpha}}{2m_{W^{(0)}}}h^{(0)}
\Big( \bar{u}^{(k)}_{1,\alpha}u^{(k)}_{1,\alpha}
+\bar{u}^{(k)}_{2,\alpha}u^{(k)}_{2,\alpha}\Big)
\nonumber \\ &&\hspace{0.5cm}
-g\cos\big( 2\theta_{u^{(k)}_\alpha} \big)
\frac{m_{u^{(0)}_\alpha}}{2m_{W^{(0)}}}h^{(0)}
\bar{u}^{(k)}_{1,\alpha}\gamma_5u^{(k)}_{2,\alpha}
+{\rm H.c.},
\label{hukuk}
\end{eqnarray}
\begin{eqnarray}
&&
{\cal L}_{h^{(0)}d^{(k)}_\alpha d^{(k)}_\alpha}=
\nonumber \\ && \hspace{0.5cm} 
-g\sin(2\theta_{d^{(k)}_\alpha})\frac{m_{d^{(0)}_\alpha}}{2m_{W^{(0)}}}h^{(0)}
\Big( \bar{d}^{(k)}_{1,\alpha}d^{(k)}_{1,\alpha}
+\bar{d}^{(k)}_{2,\alpha}d^{(k)}_{2,\alpha}\Big)
\nonumber \\ &&\hspace{0.5cm}
-g\cos\big( 2\theta_{d^{(k)}_\alpha} \big)
\frac{m_{d^{(0)}_\alpha}}{2m_{W^{(0)}}}h^{(0)}
\bar{d}^{(k)}_{1,\alpha}\gamma_5d^{(k)}_{2,\alpha}
+{\rm H.c.},
\label{hdkdk}
\end{eqnarray}
\begin{eqnarray}
{\cal L}_{u^{(0)}_\beta d^{(k)}_\alpha W^{(k)}}=\frac{g\,\kappa_{\beta\alpha}}{\sqrt{2}}W^{(k)+}_\mu\bar{u}^{(0)}_\beta\gamma^\mu P_L
\Big[\sin\theta_{d^{(k)}_\alpha}d^{(k)}_{1,\alpha}
\nonumber \\
-\cos\theta_{d^{(k)}_\alpha}d^{(k)}_{2,\alpha} 
\Big]
+{\rm H.c.},
\label{udkWk}
\end{eqnarray}
\begin{eqnarray}
{\cal L}_{d^{(0)}_\alpha u^{(k)}_\beta W^{(k)}}=\frac{g\,\kappa^*_{\beta\alpha}}{\sqrt{2}}W^{(k)-}_\mu\bar{d}^{(0)}_\alpha\gamma^\mu P_L
\Big[\sin\theta_{u^{(k)}_\beta}u^{(k)}_{1,\beta}
\nonumber \\
-\cos\theta_{u^{(k)}_\beta}u^{(k)}_{2,\beta} 
\Big]
+{\rm H.c.},
\label{dukWk}
\end{eqnarray}
\begin{eqnarray}
&&
{\cal L}_{u^{(0)}_\beta d^{(k)}_\alpha G_W^{(k)}}=\frac{ig\,\kappa_{\beta\alpha}}{\sqrt{2}\,m_{W^{(k)}}}G^{(k)+}_W\bar{u}^{(0)}_\beta 
\nonumber \\ &&\hspace{0.2cm}\times
(m_{d^{(k)}_\alpha}P_R-m_{u^{(0)}_\beta}P_L)
\Big[\sin\theta_{d^{(k)}_\alpha}d^{(k)}_{1,\alpha}
-\cos\theta_{d^{(k)}_\alpha}d^{(k)}_{2,\alpha}
\Big]
\nonumber \\ &&
\hspace{0.2cm}+{\rm H.c.},
\label{udkGwk}
\end{eqnarray}
\begin{eqnarray}
&&
{\cal L}_{d^{(0)}_\alpha u^{(k)}_\beta G_W^{(k)}}=\frac{ig\,\kappa^*_{\beta\alpha}}{\sqrt{2}\,m_{W^{(k)}}}G^{(k)-}_W\bar{d}^{(0)}_\alpha 
\nonumber \\ && \hspace{0.2cm}\times
(m_{u^{(k)}_\beta}P_R-m_{d^{(0)}_\alpha}P_L)
\Big[\sin\theta_{u^{(k)}_\beta}u^{(k)}_{1,\beta}
-\cos\theta_{u^{(k)}_\beta}u^{(k)}_{2,\beta}
\Big]
\nonumber \\ &&\hspace{0.2cm}
+{\rm H.c.},
\label{dukGwk}
\end{eqnarray}
\begin{eqnarray}
&&
{\cal L}_{u^{(0)}_\beta d^{(k)}_\alpha W^{(k)}_{\rm s}}=\frac{ig\,m_{(k)}\kappa_{\beta\alpha}}{\sqrt{2}\,m_{W^{(0)}}m_{W^{(k)}}}W^{(k)+}\bar{u}^{(0)}_\beta
\nonumber \\ && \hspace{0.5cm}\times
\Big[
\sin\theta_{d^{(k)}_\alpha}\Big( \Big(\,\frac{m^2_{W^{(k)}}}{m_{(k)}}-m_{d^{(k)}_\alpha}\Big)P_R+m_{u^{(0)}_\beta}P_L \Big)d^{(k)}_{1,\alpha}
\nonumber \\ &&  \hspace{0.5cm}
+\cos\theta_{d^{(k)}_\alpha}\Big( \Big(\,\frac{m^2_{W^{(k)}}}{m_{(k)}}+m_{d^{(k)}_\alpha}\Big)P_R-m_{u^{(0)}_\beta }P_L \Big)d^{(k)}_{2,\alpha}
\Big]
\nonumber \\ &&  \hspace{0.5cm}
+{\rm H.c.},
\label{udkWsk}
\end{eqnarray}
\begin{eqnarray}
&&
{\cal L}_{d^{(0)}_\alpha u^{(k)}_\beta W^{(k)}_{\rm s}}=\frac{ig\,m_{(k)}\kappa^*_{\beta\alpha}}{\sqrt{2}\,m_{W^{(0)}}m_{W^{(k)}}}W^{(k)-}\bar{d}^{(0)}_\alpha
\nonumber \\ &&\hspace{0.5cm} \times
\Big[
\sin\theta_{u^{(k)}_\beta}\Big( \Big( \frac{m^2_{W^{(k)}}}{m_{(k)}}-m_{u^{(k)}_\beta} \Big)P_R+m_{d^{(0)}_\alpha}P_L \Big)u^{(k)}_{1,\beta}
\nonumber \\ &&\hspace{0.5cm}
+\cos\theta_{u^{(k)}_\beta} \Big( \Big( \frac{m^2_{W^{(k)}}}{m_{(k)}}+m_{u^{(k)}_\beta} \Big)P_R-m_{d^{(0)}_\alpha}P_L \Big)u^{(k)}_{2,\beta}
\Big]
\nonumber \\ &&\hspace{0.5cm}
+{\rm H.c.}
\label{dukWsk}
\end{eqnarray}
Here, $P_L=({\bf 1}-\gamma_5)/2$ and $P_R=({\bf 1}+\gamma_5)/2$ are the {\it chiral projection operators}. Furthermore, $\kappa_{\beta\alpha}$ denotes the entries of the {\it Cabibbo-Kobayashi-Maskawa} (CKM) {\it mixing matrix}~\cite{PDG,NCabibbo,KoMa,Wolfenstein,ChKe}, with $\beta=u,c,t$ and $\alpha=d,s,b$. The CKM matrix is the only source of $CP$ violation in the 4DSM, which important for baryon asymmetry in the universe to be explained~\cite{Sakharov}. These expressions also involve the mixing angles $\theta_{q^{(k)}_\alpha}=\tan^{-1}[(m_{q^{(k)}_\alpha}+m_{(k)})/(m_{q^{(k)}_\alpha}-m_{(k)})]^{1/2}$, where $q=u,d$. \\


\section{Calculation of one-loop effects from Kaluza-Klein modes on $h^{(0)}\to q^{(0)}_\alpha q^{(0)}_\beta$}
\label{analytical}
In this section, an analytical calculation of the branching ratio for the decay $h^{(0)}\to q^{(0)}_\alpha q^{(0)}_\beta$, with $\alpha\ne\beta$, is carried out, in the context of the KK theory previously discussed. Calculations and estimations of the 4DSM contributions to this process, occurring for the first time at one loop, have been performed in Refs.~\cite{BGLMT,GAMRT}. For the present investigation, such a calculation is reviewed, together with the calculation of KK-excited-modes contributions, in order to determine the total contribution from the whole KK theory. \\

\begin{figure}[ht]
\center
\includegraphics[width=3.9cm]{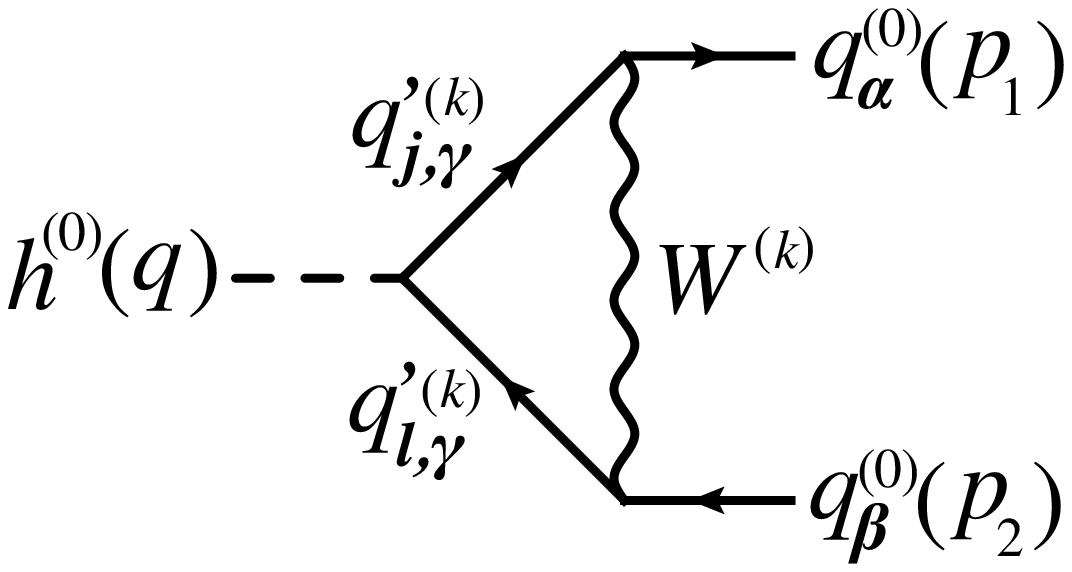}
\hspace{0.6cm}
\includegraphics[width=3.9cm]{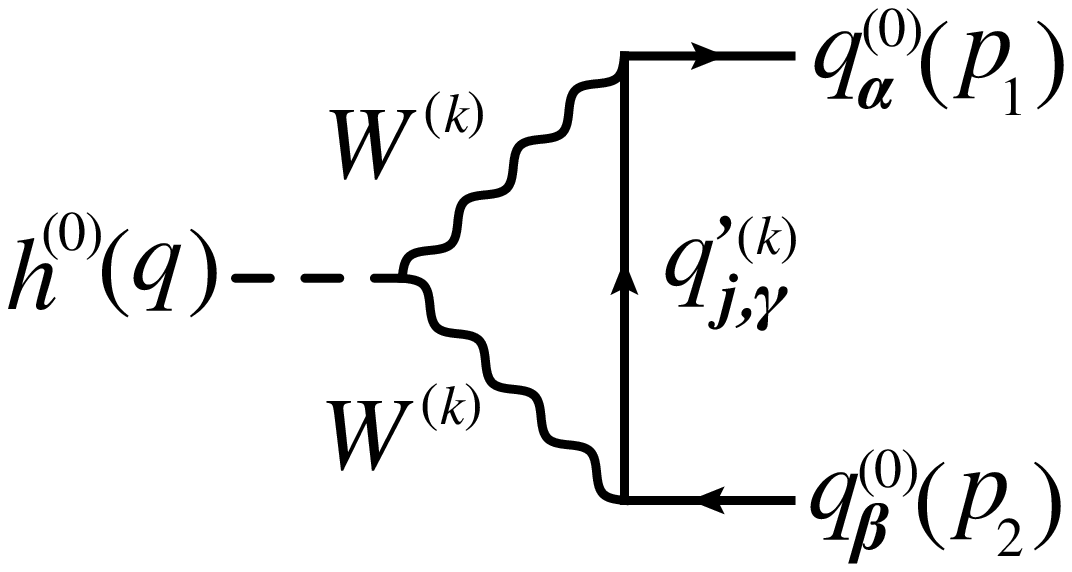}
\vspace{0.5cm}
\\
\includegraphics[width=3.9cm]{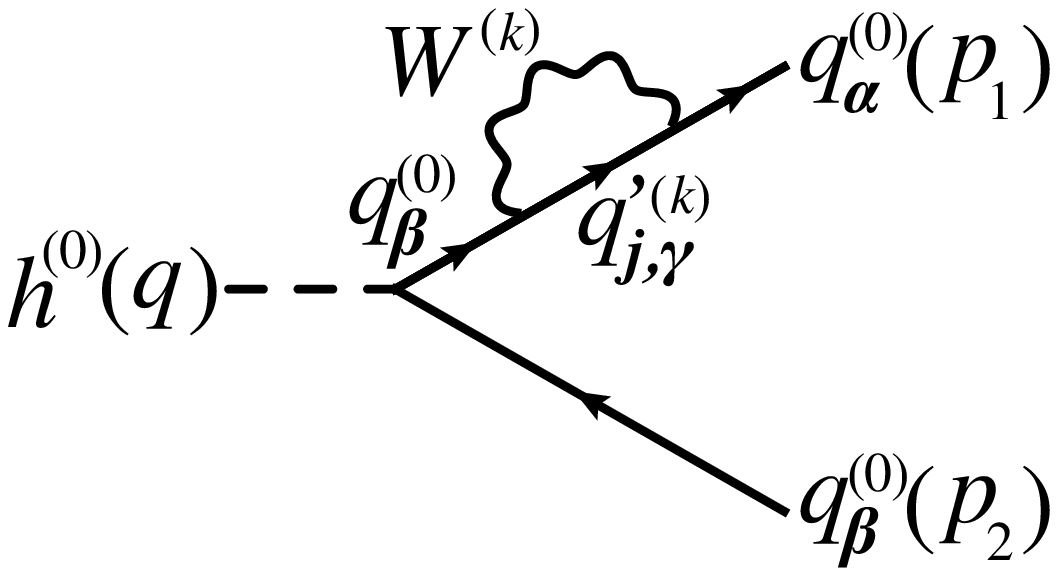}
\hspace{0.6cm}
\includegraphics[width=3.9cm]{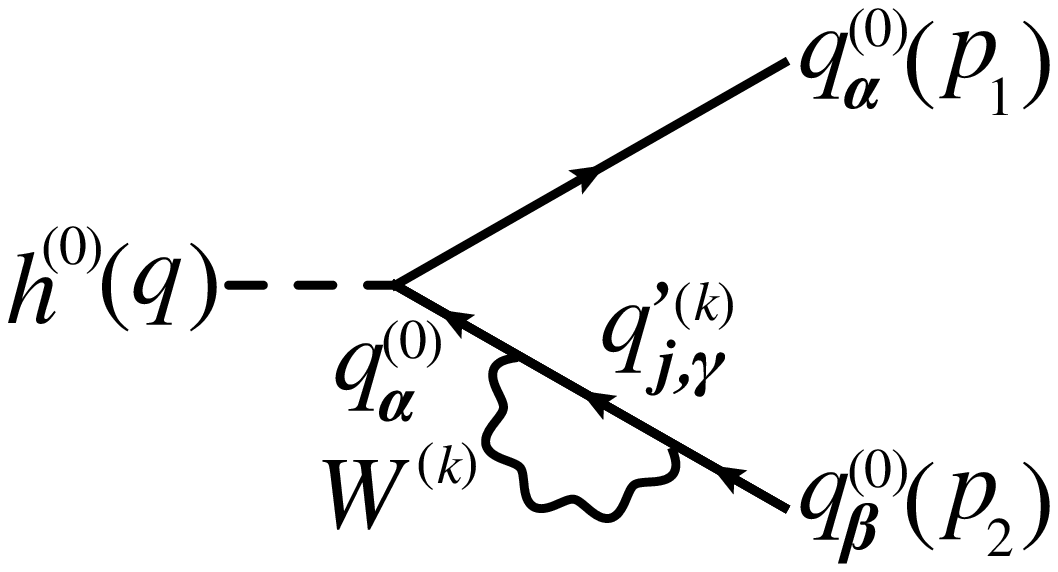}
\caption{\label{gaugediags} One-loop Feynman diagrams with virtual KK vector-boson lines contributing to $h^{(0)}\to q^{(0)}_\alpha q^{(0)}_\beta$. The presence of unprimed and primed quark fields indicates that if $q^{(0)}_\alpha$ is $u$ type, then $q'^{(k)}_{j,\gamma}$ is $d$ type, and {\it vice versa}. Moreover, $j,l=1,2$ label the mass-degenerate KK-excited-mode quark fields $q'^{(k)}_{1,\gamma}$ and $q'^{(k)}_{2,\gamma}$, whereas $\gamma$ denotes quark flavors.}
\end{figure}

\begin{figure}[ht]
\center
\includegraphics[width=3.9cm]{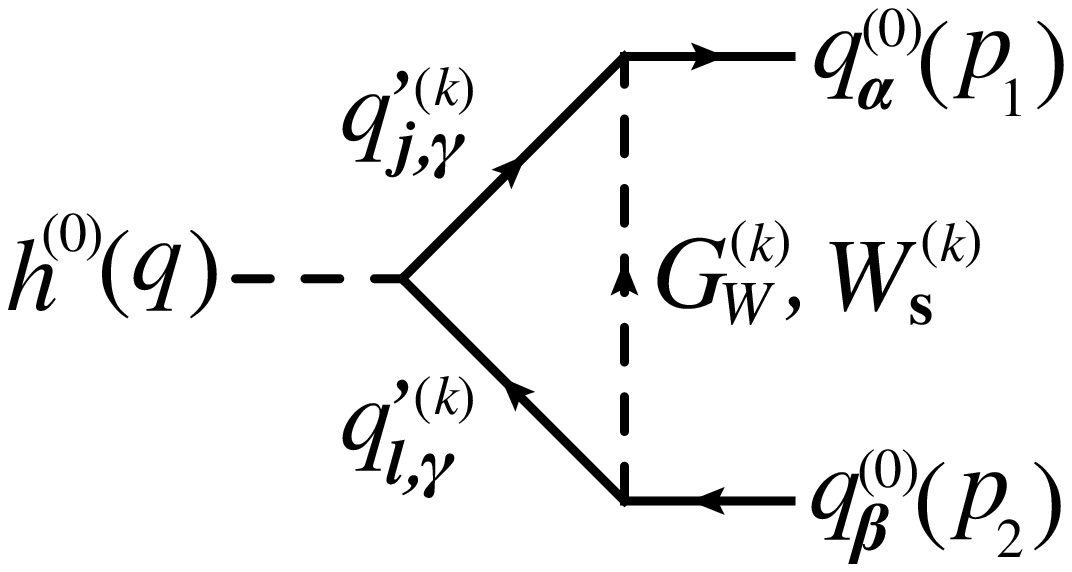}
\hspace{0.6cm}
\includegraphics[width=3.9cm]{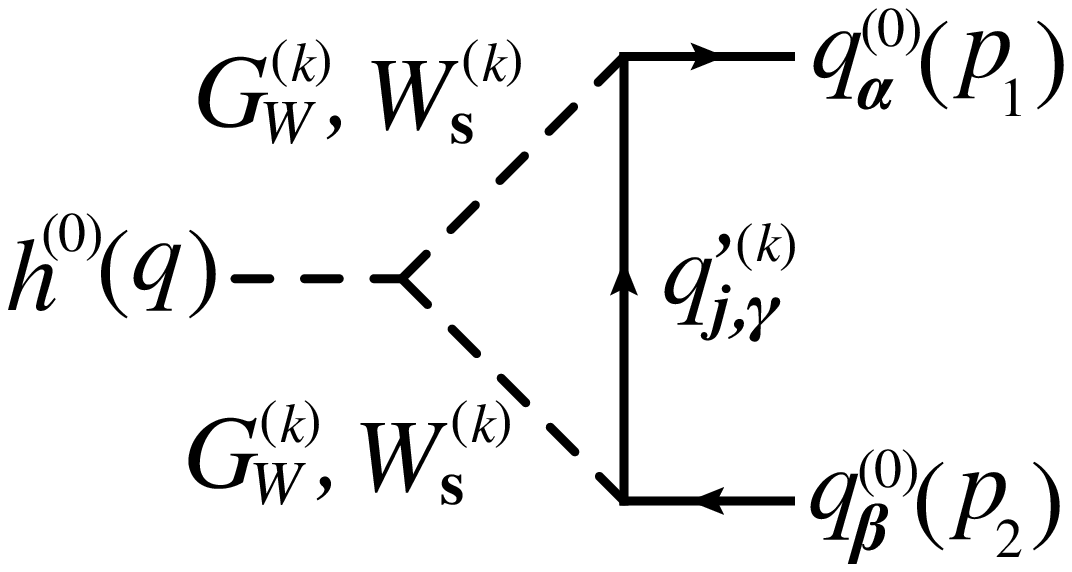}
\vspace{0.5cm}
\\
\includegraphics[width=3.9cm]{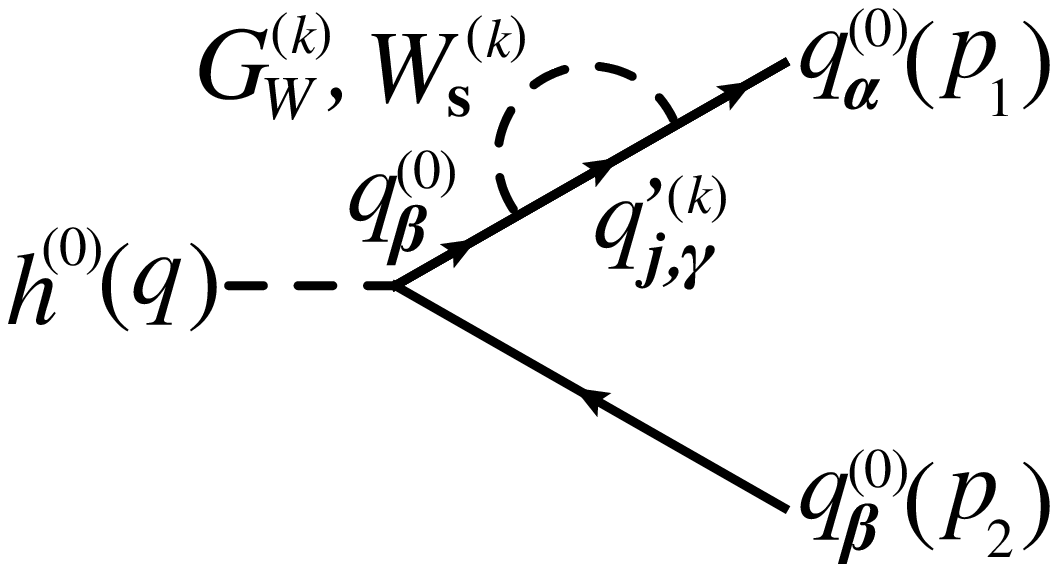}
\hspace{0.6cm}
\includegraphics[width=3.9cm]{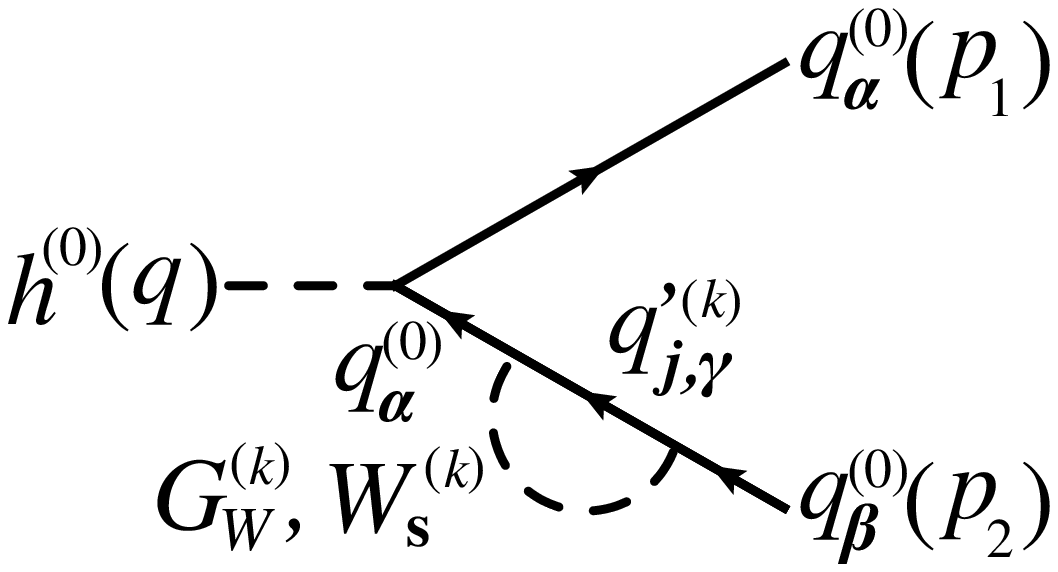}
\caption{\label{scalardiags} One-loop Feynman diagrams with virtual KK scalar-field lines contributing to $h^{(0)}\to q^{(0)}_\alpha q^{(0)}_\beta$. Such scalar lines correspond to either a pseudo-Goldstone boson $G^{(k)}_W$ or a physical scalar $W^{(k)}_{\rm s}$. The presence of unprimed and primed quark fields indicates that if $q^{(0)}_\alpha$ is $u$ type, then $q'^{(k)}_{j,\gamma}$ is $d$ type, and {\it vice versa}. Moreover, $j,l=1,2$ label the mass-degenerate KK-excited-mode quark fields $q'^{(k)}_{1,\gamma}$ and $q'^{(k)}_{2,\gamma}$, whereas $\gamma$ denotes quark flavors.}
\end{figure}

\begin{figure}[ht]
\center
\includegraphics[width=3.9cm]{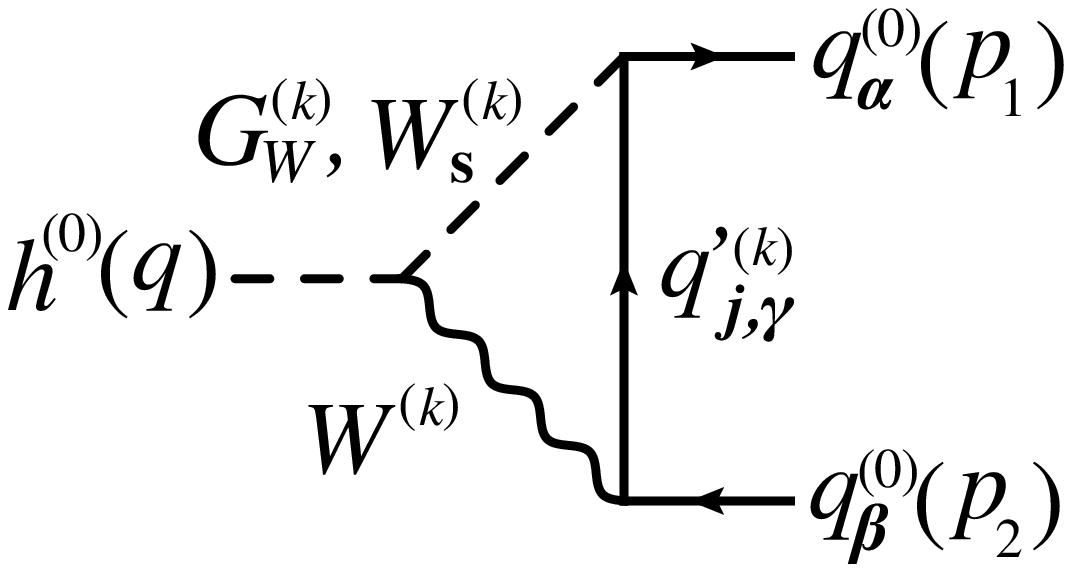}
\hspace{0.6cm}
\includegraphics[width=3.9cm]{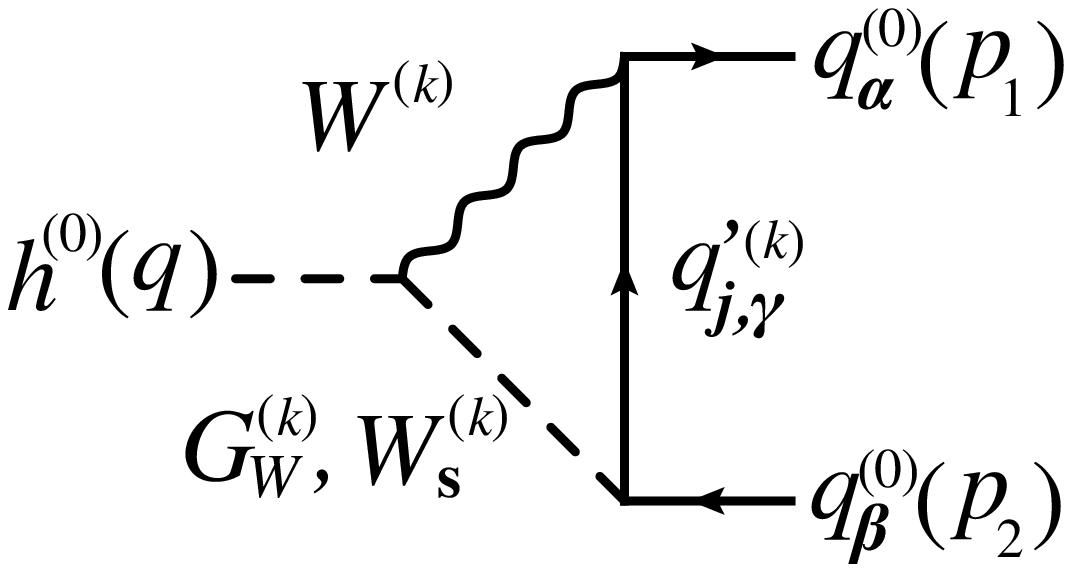}
\vspace{0.5cm}
\\
\includegraphics[width=3.9cm]{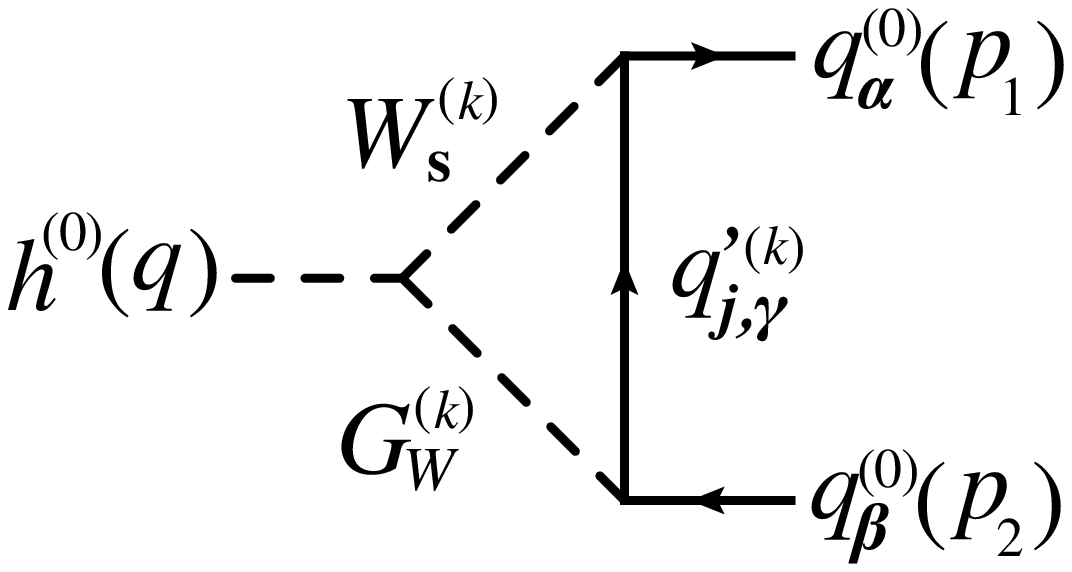}
\hspace{0.6cm}
\includegraphics[width=3.9cm]{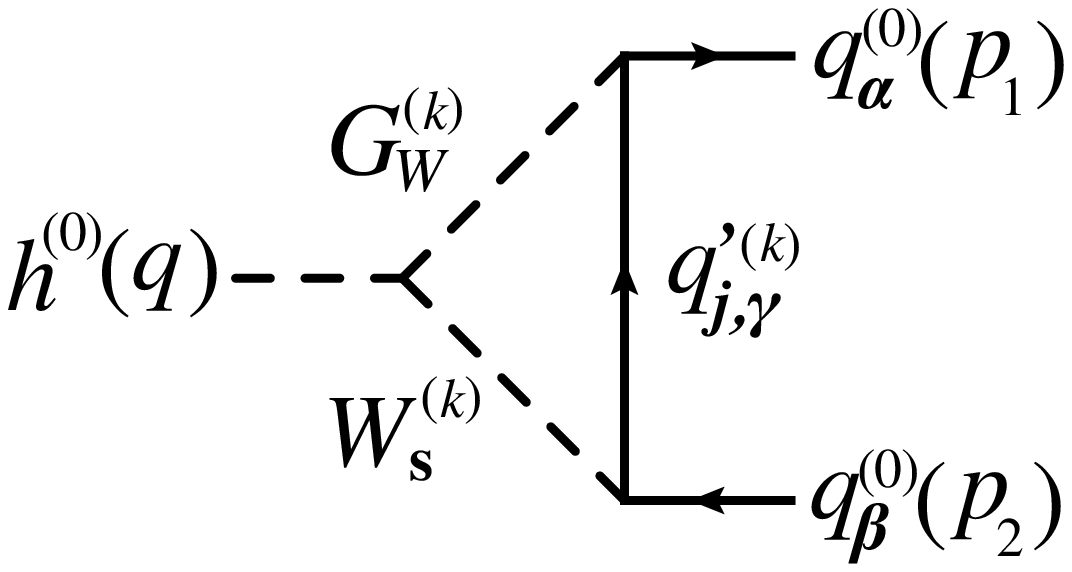}
\caption{\label{combdiags} One-loop Feynman diagrams with both virtual KK vector- and scalar-field lines contributing to $h^{(0)}\to q^{(0)}_\alpha q^{(0)}_\beta$. Scalar-field lines correspond to either a pseudo-Goldstone boson $G^{(k)}_W$ or a physical scalar $W^{(k)}_{\rm s}$. The presence of unprimed and primed quark fields indicates that if $q^{(0)}_\alpha$ is $u$ type, then $q'^{(k)}_{j,\gamma}$ is $d$ type, and {\it vice versa}. Moreover, $j,l=1,2$ label the mass-degenerate KK-excited-mode quark fields $q'^{(k)}_{1,\gamma}$ and $q'^{(k)}_{2,\gamma}$, whereas $\gamma$ denotes quark flavors.}
\end{figure}

The Feynman diagrams characterizing this decay process are shown in Figs.~\ref{gaugediags}-\ref{combdiags}.
All such diagrams include virtual KK-quark-field lines, which have been generically denoted as $q'^{(k)}_{j,\gamma}$, with $k$ a non-negative integer, whereas $q^{(0)}_\alpha$ and $q^{(0)}_\beta$ represent external quark fields. Since in this model quark-flavor change takes place through quark mixing in charged currents, diagrams in which the external quarks are $u$ type have $d$-type virtual KK quark lines. On the other hand, if the external fermion lines correspond to $d$ quarks, then the internal quark lines are associated to $u$-type KK-excited-mode quarks. 
Primed virtual-quark labels $q'^{(k)}_{j,\gamma}$ and unprimed external quarks labels $q^{(0)}_\alpha$ in the diagrams have been utilized to characterize such a distinction. Virtual-field lines in the diagrams of Fig.~\ref{gaugediags} are exclusively associated to KK-excited-mode quarks $q'^{(k)}_{j,\gamma}$ and vector fields $W^{(k)\pm}_\mu$. Virtual lines in the diagrams of Fig.~\ref{scalardiags} correspond only to virtual KK quark fields $q'^{(k)}_{j,\gamma}$ and to either KK pseudo-Goldstone bosons $G_W^{(k)\pm}$ or physical scalars $W^{(k)\pm}$. Scalar-field-changing vertices are considered in Fig.~\ref{combdiags}, which also includes diagrams with combining vertices $h^{(0)}W^{(k)}_\mu G^{(k)}_W$ or $h^{(0)}W^{(k)}_\mu W^{(k)}$.  The 4DSM includes no physical charged scalar fields, so diagrams involving physical zero-mode charged scalars are absent. Furthermore, recall that we have chosen the unitary gauge to remove zero-modes gauge symmetry, in which case the calculation of the 4DSM contributions includes no diagrams with zero-mode pseudo-Goldstone bosons, thus leaving only zero-mode contributions from the gauge diagrams given in Fig.~\ref{gaugediags}.\\

As for the contributing Feynman diagrams displayed in Fig.~\ref{gaugediags}, consider, for a moment, the triangle diagram with two KK quark-field virtual lines. For fixed KK index $(k)$ and quark flavor $\gamma$, it represents four cases, defined by the index values $j,l=1,2$. The two diagrams with $l\ne j$ coincide in every aspect but a differing global sign, which yields an exact cancellation among them. Conversely, if $l=j$, the resulting analytical expressions for the two corresponding diagrams are the same, except for global factors $\cos^2\theta_{q^{(k)}_\alpha}$ and $\sin^2\theta_{q^{(k)}_\alpha}$, which lead, through trigonometric identities, to the cancellation of the $\theta_{q^{(k)}_\alpha}$ dependence from the sum of such contributions. Analogous cancellations of the $\theta_{q^{(k)}_\alpha}$ dependence replicate for all other diagrams in Fig.~\ref{gaugediags} when considering the cases $j=1,2$ and adding the resulting contributions together. Qualitatively, the same thing happens with the combinations of those diagrams of Fig.~\ref{scalardiags} in which virtual pseudo-Goldstone-boson lines are involved. Contrariwise, such convenient cancellations and combinations do not occur for the diagrams of Fig.~\ref{scalardiags} with physical-scalar lines. \\

The sum of the whole set of Feynman diagrams given in Figs.~\ref{gaugediags}-\ref{combdiags} produces the amplitude $i{\cal M}=\bar{u}_\alpha(p_1)\,\Gamma^{\alpha\beta}(q)\,v_\beta(p_2)$, where $\Gamma^{\alpha\beta}$ has the structure
\begin{equation}
\Gamma^{\alpha\beta}=\sum_{k=0}^\infty\sum_\gamma{\cal B}_{\alpha\gamma}\,{\cal B}^*_{\beta\gamma}\Big( h^{(k)\gamma}_{1,\alpha\beta}\cdot{\bf 1}_4+h^{(k)\gamma}_{2,\alpha\beta}\,\gamma_5 \Big),
\label{Mamplitude}
\end{equation}
with ${\bf 1}_4$ denoting the $4\times4$ identity matrix in Dirac-matrix space. In this expression, $\gamma=d,s,b$ if final-state fermions are $u$-type quarks, whereas for $d$-type external quarks such an index runs over $\gamma=u,c,t$. Moreover, the factors ${\cal B}_{\alpha\beta}$ relate to CKM matrix elements, $\kappa_{\alpha\beta}$, as
\begin{equation}
{\cal B}_{\alpha\beta}=
\left\{
\begin{array}{l}
\kappa_{\alpha\beta}, \,\,\textrm{if $q^{(0)}$ are $u$-type},
\vspace{0.3cm}
\\
\kappa^*_{\beta\alpha}, \,\,\textrm{if $q^{(0)}$ are $d$-type}.
\end{array}
\right.
\end{equation}
The calculation of these contributions has been accomplished by following the {\it dimensional regularization} approach~\cite{BoGi} and the {\it Passarino-Veltman tensor reduction method}~\cite{PassVe}, implemented through the software {\it Mahtematica}, by Wolfram, and the package {\it Feyncalc}~\cite{MBD}. Explicit expressions of the coefficients $h^{(k)\gamma}_{1,\alpha\beta}$ and $h^{(k)\gamma}_{2,\alpha\beta}$, which charcterize Eq.~(\ref{Mamplitude}), are provided for the case $(k)\ne(0)$ in Appendix~\ref{App}. We have verified that UV divergences proceeding from loop integrals in the coefficients $h^{(k)\gamma}_{j,\alpha\beta}$ vanish exactly. In general, the specific manner in which the elimination of such UV divergences takes place, for some given calculation, depends on the gauge choice. For instance, these cancellations are usually more intricate in the unitary gauge, in which gauge propagators increase the {\it superficial degree of divergence} of loop integrals. In the nonlinear gauge-fixing approach, followed throughout the present investigation to calculate the contributions from KK excited modes, expressions emerged from the diagrams of Fig.~\ref{gaugediags} are finite by themselves, upon operation of the {\it Glashow-Iliopoulos-Maiani mechanism}~\cite{GIMmech}. And the same holds for diagrams with virtual pseudo-Goldstone-boson lines in Figs.~\ref{scalardiags} and \ref{combdiags}. Furthermore, if the whole set of contributing diagrams are taken at once, the resulting expression is finite, even without intervention of such a mechanism. \\

The afore-described cancellation of divergences exclusively regards UV divergences originating in loop integrals, which are continuous sums over four-dimensional momenta. In the general context of a field theory defined in a $(4+n)$-dimensional spacetime, the contributions from KK excited modes to standard Green's functions\footnote{The term {\it standard Green's function} refers to any Green's function with all its external lines corresponding to KK zero modes.} include multiple infinite sums running over discrete extra-dimensional momenta, namely the KK sums, which are a source of further divergences. This is expected on the grounds of Dyson's criterion of renormalizability, according to which extra-dimensional field theories are nonrenormalizable~\cite{PeSch}. The KK theory of the Standard Model in $4+n$ dimensions comprises only four-dimensional couplings which, from the same criterion, are renormalizable, thus suggesting that the presence of divergent KK sums is a manifestation of nonrenormalizability inherited from the original theory in $4+n$ dimensions. It has been argued, in Refs.~\cite{MNNST,HNT,LMNTo,GMNNTT}, that such discrete divergences are genuine UV divergences, though associated to short-distance effects in the compact extra dimensions, as they come from sums that include infinitely-large values of extra-dimensional momenta. Since for the present investigation only one extra dimension has been assumed to exist, no such KK divergences arise in our phenomenological calculation, so elimination of divergences from loop integrals suffices to ensure UV finiteness of the KK contributions. This statement is better appreciated in a scenario of large compactification scale $R^{-1}$, where the expressions can be expanded in powers of $(R/k)^2$, which directly turn KK sums into finite {\it Riemann zeta functions}, $\zeta(2s)=\sum_{k=1}^\infty k^{-2s}$, with $s$ a natural number so that KK contributions decouple in the limit as $R^{-1}\to\infty$. Finiteness of exact contributions, with no regard to $R$-series, has been recently discussed in detail for five-dimensional {\it quantum electrodynamics}~\cite{MNNST}, five-dimensional {\it Yang-Mills theories}~\cite{HNT}, $\lambda\phi^4$ theory~\cite{LMNTo} in five spacetime dimensions, and the 5DSM~\cite{GMNNTT}, where an intricate interplay between the one-dimensional {\it Epstein function}~\cite{Epstein}, emanated from KK sums, and the {\it gamma function}, proceeding from dimensional regularization, has been shown to produce nontrivial eliminations of KK-sums divergences. The presence of two or more extra dimensions complicates things, since in such cases amplitudes carry multiple KK sums, thus yielding divergent results, with divergences characterized by poles of the Epstein function~\cite{GNT,MNNST,HNT,LMNTo,GMNNTT}. In such cases, renormalization has to be implemented in a modern sense~\cite{Weinbergbook}, supported by the formalism of effective Lagrangians. Being nonrenormalizable, field theories of extra dimensions cannot be fundamental and thus must include an infinite sum of Lagrangian terms with increasing canonical dimensions, which involve a higher-energy scale to be interpreted as the characteristic scale of some fundamental physical description of nature~\cite{Wudka}, beyond extra dimensions. The Lagrangian terms with canonical dimensions greater than $4+n$ are extra-dimensional analogues of the nonrenormalizable interactions that constitute effective theories in four dimensions~\cite{BuWy,LLR}. After compactification, effective-Lagrangian interactions with canonical dimensions greater than $4+n$ generate the necessary counterterms to carry out renormalization. According to the formalism of effective theories, such terms are understood to be a parametrization of effects produced at lower energies by the fundamental higher-energy formulation~\cite{Wudka}. In this sense, UV divergences from KK sums are absorbed by the fundamental higher-energy physical description. These ideas have been implemented and thoroughly discussed, in KK theories, in Refs.~\cite{MNNST,HNT,LMNTo,GMNNTT}. \\

From Eq.~(\ref{Mamplitude}), the {\it decay rate} for the process $h^{(0)}\to \bar{q}^{(0)}_\alpha q^{(0)}_\beta+\bar{q}^{(0)}_\beta q^{(0)}_\alpha$, briefly referred to as $h^{(0)}\to q^{(0)}_\alpha q^{(0)}_\beta$, is split into a sum of three terms as $\Gamma(h^{(0)}\to q^{(0)}_\alpha q^{(0)}_\beta)=\Gamma_{\rm SM}+\Gamma_{\rm SMKK}+\Gamma_{\rm KK}$, with $\Gamma_{\rm SM}$ the 4DSM contribution, $\Gamma_{\rm SMKK}$ the interference contribution produced by KK zero and excited modes, and where $\Gamma_{\rm KK}$ is the contribution exclusively generated by KK excited modes. By defining
\begin{equation}
h^{\rm SM}_{j,\alpha\beta}=\sum_{m=2,3}{\cal B}_{\alpha\gamma_m}\,{\cal B}^*_{\beta\gamma_m} \big(h^{(0)\gamma_m}_{j,\alpha\beta}-h^{(0)\gamma_1}_{j,\alpha\beta}\big),
\label{hSM}
\end{equation}
\begin{equation}
h^{\rm KK}_{j,\alpha\beta}=\sum_{k=1}^\infty\sum_{m=2,3}{\cal B}_{\alpha\gamma_m}\,{\cal B}^*_{\beta\gamma_m} \big(h^{(k)\gamma_m}_{j,\alpha\beta}-h^{(k)\gamma_1}_{j,\alpha\beta}\big),
\label{hKK}
\end{equation}
the aforementioned decay-rate terms are expressed as
\begin{equation}
\Gamma_{\rm SM}=N_C\frac{\sqrt{M^{(0)}_+M^{(0)}_-}}{4\pi m^3_{h^{(0)}}}\big( |h^{\rm SM}_{1,\alpha\beta}|^2M^{(0)}_++|h^{\rm SM}_{2,\alpha\beta}|^2M^{(0)}_- \big),
\label{drateSM}
\end{equation}
\begin{eqnarray}
&&
\Gamma_{\rm SMKK}=N_C\frac{\sqrt{M^{(0)}_+M^{(0)}_-}}{2\pi m^3_{h^{(0)}}}\Big( {\rm Re}\{ h^{{\rm SM}*}_{1,\alpha\beta}\,\,h^{\rm KK}_{1,\alpha\beta} \}M^{(0)}_+
\nonumber \\ &&
\hspace{3.1cm}
+{\rm Re}\{ h^{{\rm SM}*}_{2,\alpha\beta}\,\,h^{\rm KK}_{2,\alpha\beta} \}M^{(0)}_- \Big),
\label{drateSMKK}
\end{eqnarray}
\begin{equation}
\Gamma_{\rm KK}=N_C\frac{\sqrt{M^{(0)}_+M^{(0)}_-}}{4\pi m^3_{h^{(0)}}}\big( |h^{\rm KK}_{1,\alpha\beta}|^2M^{(0)}_++|h^{\rm KK}_{2,\alpha\beta}|^2M^{(0)}_- \big),
\label{drateKK}
\end{equation}
with $M^{(0)}_\pm=m^2_{h^{(0)}}-(m_{q^{(0)}_\alpha}\pm m_{q^{(0)}_\beta})^2$ and $N_C=3$ the {\it color index}. In order to write down Eqs.~(\ref{hSM}) and (\ref{hKK}), the Glashow-Iliopoulos-Maiani mechanism, induced by the unitarity condition $\kappa\kappa^\dag=\kappa^\dag\kappa={\bf 1}_3$ characterizing the CKM matrix, has been implemented. These expressions comprehend two scenarios: (1) external quarks $q^{(0)}$ are $u$-type, in which case $\gamma_1=d$, $\gamma_2=s$, and $\gamma_3=b$; (2) external quarks $q^{(0)}$ are $d$-type, for which $\gamma_1=u$, $\gamma_2=c$, and $\gamma_3=t$. \\

Regarding the contributions from KK excited modes, we assume a very large compactification scale $R^{-1}$, and within such a context we express the factors $h^{\rm KK}_{j,\alpha\beta}$ as the $R$-power series
\begin{eqnarray}
&&
h^{\rm KK}_{j,\alpha\beta}=iR^2\left(\frac{\alpha\pi}{s_W^2}\right)^{\frac{3}{2}}\frac{m_{q^{(0)}_\beta}+(-1)^{j-1}m_{q^{(0)}_\alpha}}{1152\,m_{W^{(0)}}^3}
\nonumber \\ &&
\hspace{0.8cm}
\times
\sum_{m=2,3}{\cal B}_{\alpha \gamma_m}\,{\cal B}^*_{\beta \gamma_m}(m^2_{\gamma_m^{(0)}}-m^2_{\gamma_1^{(0)}}) \big(10m_{\gamma_m^{(0)}}^2
\nonumber \\ &&
\hspace{0.8cm}
+10m_{\gamma_1^{(0)}}^2
+2m_{h^{(0)}}^2-2m_{q^{(0)}_\alpha}^2
+(-1)^jm_{q^{(0)}_\alpha}m_{q^{(0)}_\beta}
\nonumber \\ &&
\hspace{0.8cm}
-2m_{q^{(0)}_\beta}^2
+2m_{W^{(0)}}^2\big)+{\cal O}(R^4),
\label{serieshjKK}
\end{eqnarray}
where $\alpha$ is the fine-structure constant. In accordance with the {\it Appelquist-Carazzone decoupling theorem}~\cite{Wudka,AppCar}, decoupling of extra-dimensional effects in the limit as $R^{-1}\to\infty$ is explicit in this equation. Higher-order terms are proportional to even powers of the compactification radius $R$.


\section{Numerical estimations and discussion}
\label{numbers}
In what follows, the results found in the previous section are taken advantage of, aiming at the determination of the leading contributions from the extra-dimensional physics to the branching ratios of quark-flavor-changing Higgs decay processes $h^{(0)}\to q^{(0)}_\alpha q^{(0)}_\beta$, where the final states $b^{(0)}s^{(0)}$, $b^{(0)}d^{(0)}$, $s^{(0)}d^{(0)}$, and $c^{(0)}u^{(0)}$ are considered. Using the Higgs-boson total width $\Gamma_h=4.403\times10^{-3}\,{\rm GeV}$~\cite{PDG}, we find the following 4DSM predictions for branching ratios:
\begin{equation}
{\rm Br}(h^{(0)}\to b^{(0)}s^{(0)})_{\rm SM}=1.78\times10^{-7},
\end{equation}
\begin{equation}
{\rm Br}(h^{(0)}\to b^{(0)}d^{(0)})_{\rm SM}=8.36\times10^{-9},
\end{equation}
\begin{equation}
{\rm Br}(h^{(0)}\to s^{(0)}d^{(0)})_{\rm SM}=8.65\times10^{-15},
\end{equation}
\begin{equation}
{\rm Br}(h^{(0)}\to c^{(0)}u^{(0)})_{\rm SM}=8.13\times10^{-19}.
\end{equation}
We utilized the package {\it LoopTools}~\cite{HaPe,OldVer} to perform the numerical evaluations which leaded to these results. \\

Following the model-independent approach provided by the effective-Lagrangians technique, the authors of Ref.~\cite{BEI} investigated and discussed whether or not quark-flavor-changing interactions of the Higgs boson would be directly measurable by the LHC, estimating the upper bound ${\rm Br}(h^{(0)}\to b^{(0)}s^{(0)})_{\rm NP}<4.1\times 10^{-4}$ on the impact of new physics contributing to such processes. Using the results of that paper, the upper bounds ${\rm Br}(h^{(0)}\to b^{(0)}d^{(0)})_{\rm NP}<1.9\times10^{-5}$, ${\rm Br}(h^{(0)}\to s^{(0)}d^{(0)})_{\rm NP}<4.1\times10^{-7}$, and ${\rm Br}(h^{(0)}\to c^{(0)}u^{(0)})_{\rm NP}<2.6\times10^{-6}$, on the other quark-flavor-changing Higgs decays, are estimated. These values are well above 4DSM predictions, but lie beyond sensitivity of the LHC. Bounds on these processes, also established through the formalism of effective Lagrangians, were reported in Ref.~\cite{HKZ} as well. It is worth mentioning the quark-flavor-changing electromagnetic decays $q^{(0)}_\alpha\to A^{(0)}_\mu q^{(0)}_\beta$ of quarks, among which the top-quark decay $t^{(0)}\to A^{(0)}_\mu c^{(0)}$ is particularly interesting. The 4DSM prediction for the corresponding branching ratio is ${\rm Br}(t^{(0)}\to A^{(0)}_\mu c^{(0)})_{\rm SM}=2.31\times10^{-13}$~\cite{EHS}. Furthermore, the KK theory from the 5DSM has been shown to yield branching-ratio contributions as large as $\sim10^{-15}$~\cite{MNT}, in accordance with current lower limits on the compactification scale $R^{-1}$. \\

About bounds on $R^{-1}$, most available results regard the case of one extra dimension. In the so-called models of {\it minimal universal extra dimensions}, supersymmetry-searches data from the LHC were utilized in Ref.~\cite{DFK}  to establish the limit $1.4\,{\rm TeV}\lesssim R^{-1}$. Also from LHC data, the bound $1\,{\rm TeV}\lesssim R^{-1}$ has been estimated~\cite{BDDM}. In Ref.~\cite{BKP}, the lower bound $1.3\,{\rm TeV}\lesssim R^{-1}$ was established by investigating the contributions from KK dark matter to {\it relic density}. Data from searches of the 4DSM Higgs boson in the LHC were analyzed in Ref.~\cite{BBBKP}, yielding the less-stringent bound $0.5\,{\rm TeV}\lesssim R^{-1}$. The decay process $\bar{B}\to X_s\gamma$ has also been considered in order to bound the compactification scale, resulting in the limit $0.6\,{\rm TeV}\lesssim R^{-1}$~\cite{HaWe}. In the context of a non-minimal model of universal extra dimensions, enriched by the presence {\it boundary localized kinetic terms}~\cite{DGKN,CTW,APS}, the authors of Ref.~\cite{FKKMP} were able to give a more stringent bound on the compactification scale: $2.4\,{\rm TeV}\lesssim R^{-1}$. Recently, an investigation on the process $g^{(0)}g^{(0)}\to h^{(0)}\to\gamma^{(0)}\gamma^{(0)}$ was carried out in Ref.~\cite{GMNNTT}, from which the bounds $R^{-1}\geq 1.55\,{\rm TeV}$, $2.45\,{\rm TeV}$, $3.57\,{\rm TeV}$, $5.10\,{\rm TeV}$, and $7.25\,{\rm TeV}$  were respectively established for the Standard Model in spacetimes with 2, 4, 6, 8, and 10 universal extra dimensions, all of them compactified on $S^1/Z_2$ and all of them assumed to be the same size.\\

\begin{figure}[ht]
\center
\includegraphics[width=8.5cm]{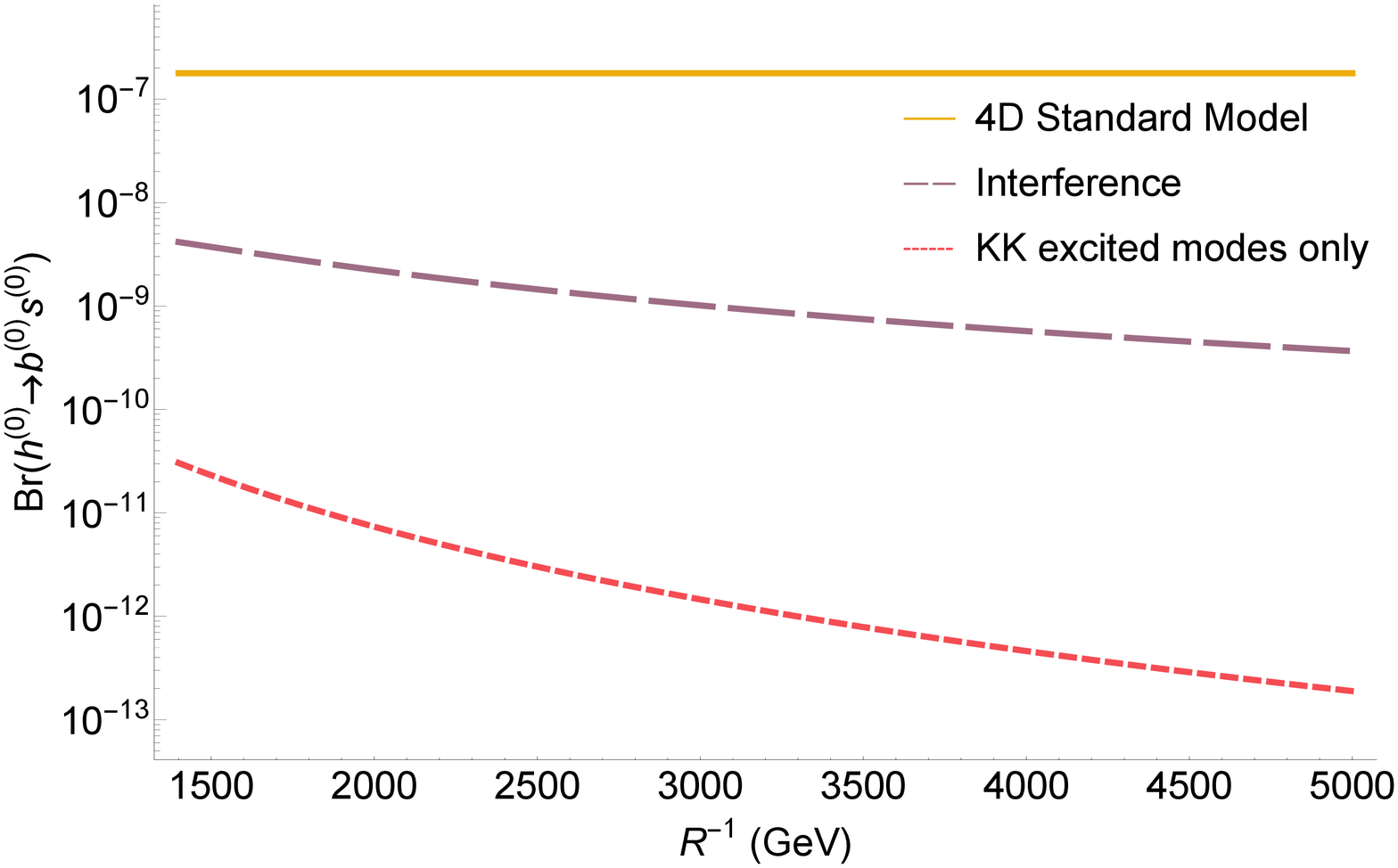}
\caption{\label{brcontrhbs1} Terms of the branching ratio of $h^{(0)}\to b^{(0)}s^{(0)}$ within $1.4\,{\rm TeV}<R^{-1}<5\,{\rm TeV}$, plotted in logarithmic scale. Such terms are the 4DSM contribution (solid horizontal line), the interference of 4DSM and extra-dimensional contributions (large-dashed curve), and the contribution from KK excited modes only (short-dashed curve).}
\end{figure}

\begin{figure}[ht]
\includegraphics[width=8.5cm]{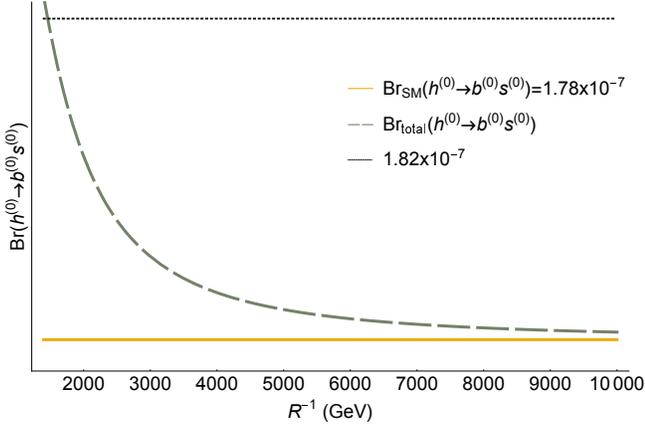}
\caption{\label{brcontrhbs2} Extra-dimensional effects on the branching ratio ${\rm Br}(h^{(0)}\to b^{(0)}s^{(0)})$. The graph includes a plot for the sole 4DSM contribution (lower horizontal line), and a plot for the total contribution from the whole KK theory (dashed curve).}
\end{figure}

\begin{figure}[ht]
\center
\includegraphics[width=8.5cm]{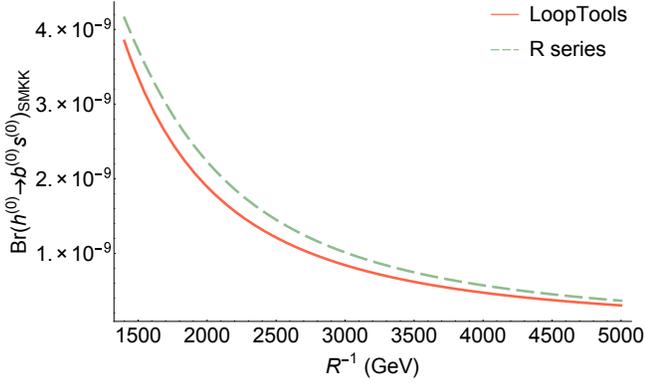}
\caption{\label{servslt} A comparison between numerical (solid curve) and $R$-series (dashed curve) evaluations of the interference term of ${\rm Br}(h^{(0)}\to b^{(0)}s^{(0)})$. Three KK-sum terms were evaluated with LoopTools for the solid curve to be plotted.}
\end{figure}

\begin{figure}[ht]
\center
\includegraphics[width=8.5cm]{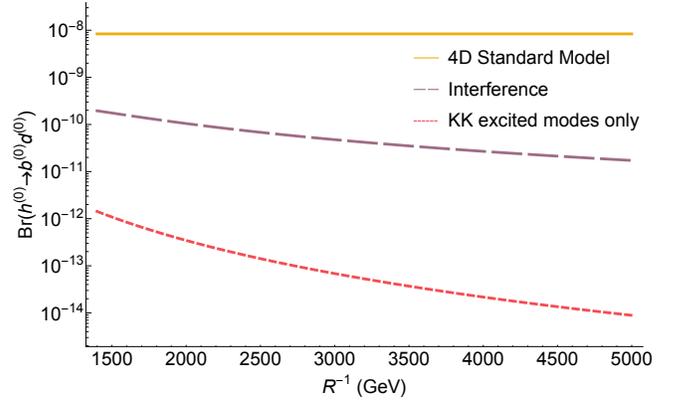}
\caption{\label{brcontrhbd1} Terms of the branching ratio of $h^{(0)}\to b^{(0)}d^{(0)}$ within $1.4\,{\rm TeV}<R^{-1}<5\,{\rm TeV}$, plotted in logarithmic scale. Such terms are the 4DSM contribution (solid horizontal line), the interference of 4DSM and extra-dimensional contributions (large-dashed curve), and the contribution from KK excited modes only (short-dashed curve).}
\end{figure}

\begin{figure}[ht]
\center
\includegraphics[width=8.5cm]{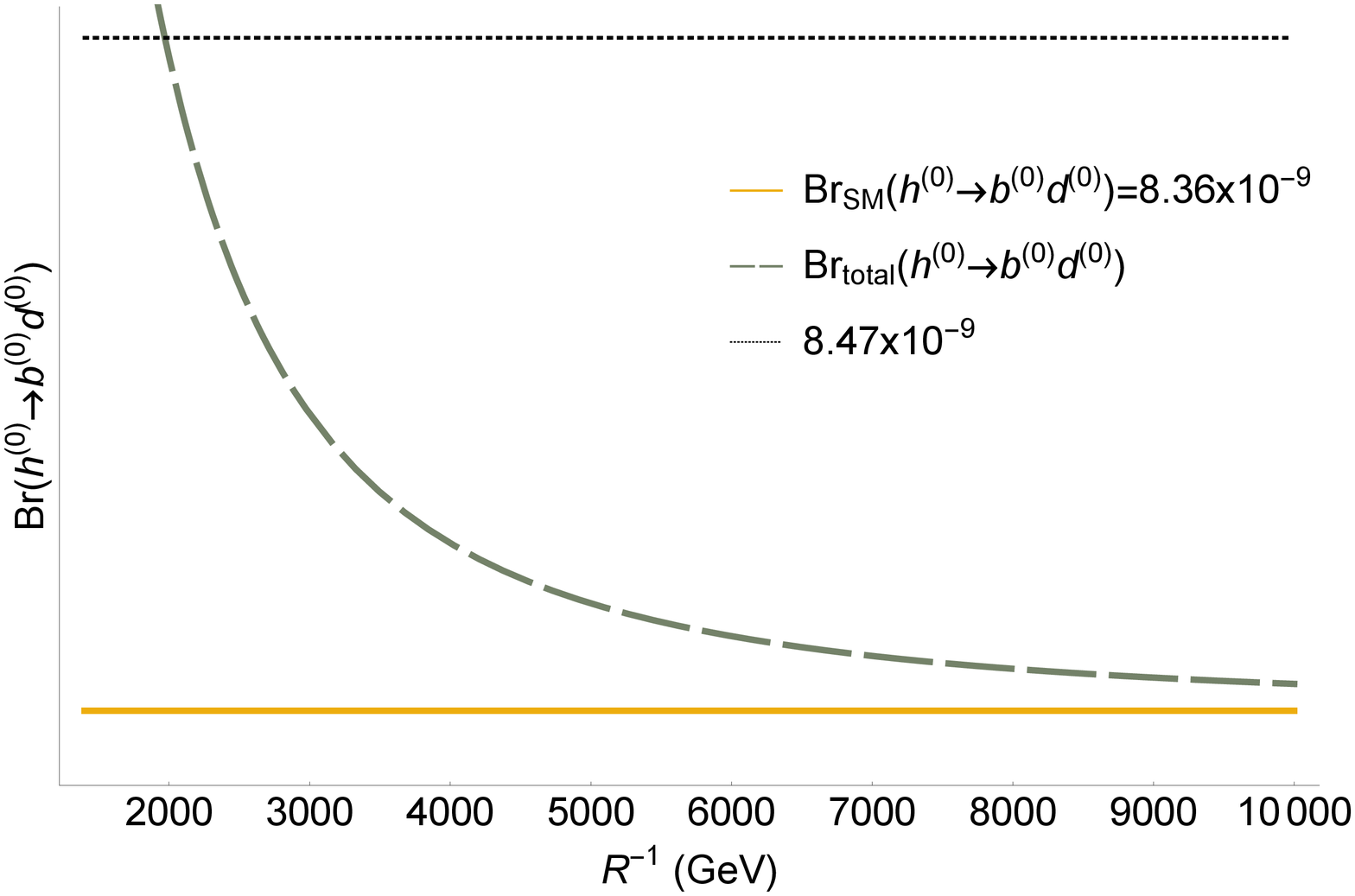}
\caption{\label{brcontrhbd2} Extra-dimensional effects on the branching ratio ${\rm Br}(h^{(0)}\to b^{(0)}d^{(0)})$. The graph includes a plot for the sole 4DSM contribution (lower horizontal line), and a plot for the total contribution from the whole KK theory (dashed curve).}
\end{figure}

\begin{figure}[ht]
\center
\includegraphics[width=8.5cm]{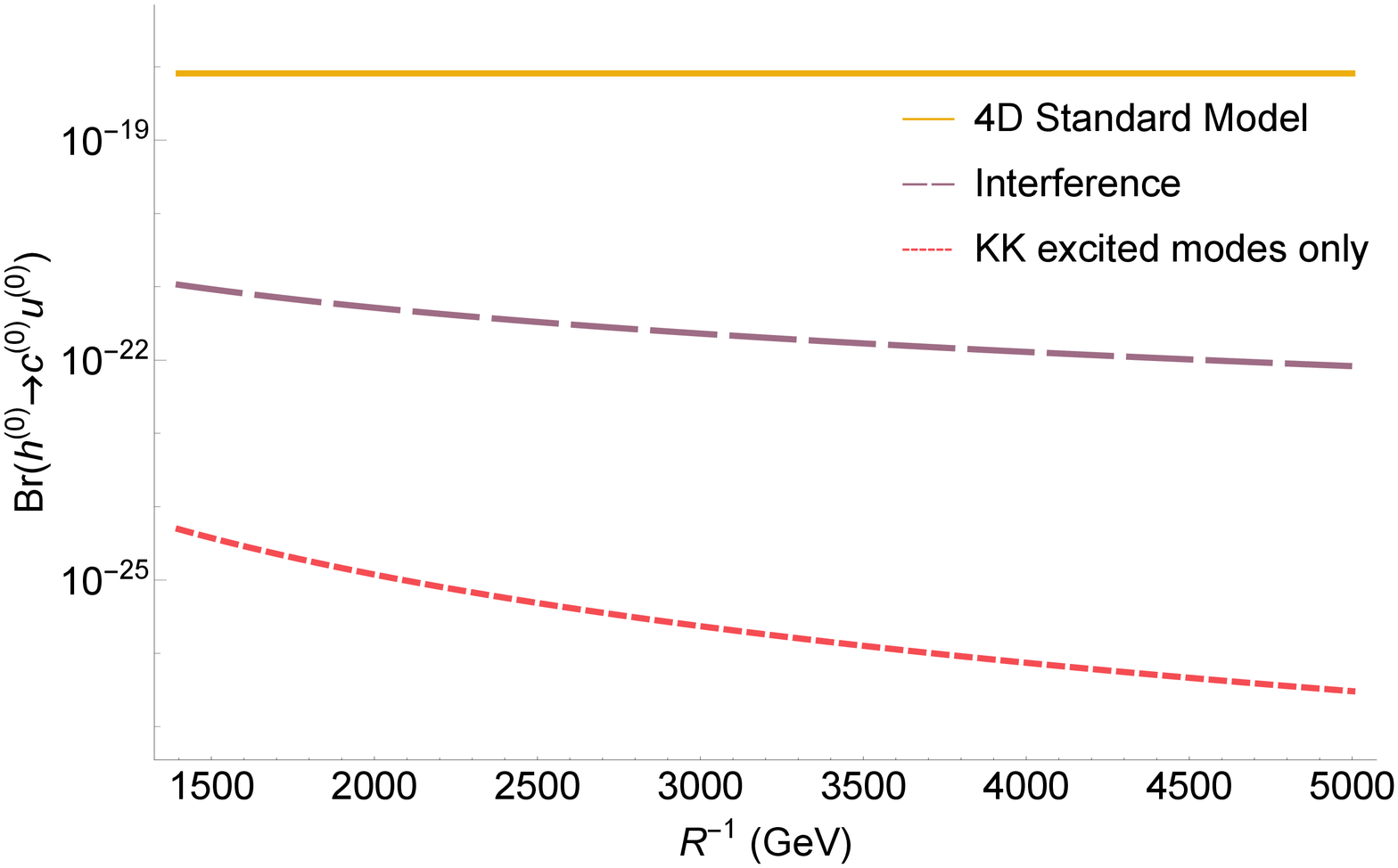}
\caption{\label{brcontrhcu1} Terms of the branching ratio of $h^{(0)}\to c^{(0)}u^{(0)}$ within $1.4\,{\rm TeV}<R^{-1}<5\,{\rm TeV}$, plotted in logarithmic scale. Such terms are the 4DSM contribution (solid horizontal line), the interference of 4DSM and extra-dimensional contributions (large-dashed curve), and the contribution from KK excited modes only (short-dashed curve).}
\end{figure}

\begin{figure}[ht]
\center
\includegraphics[width=8.5cm]{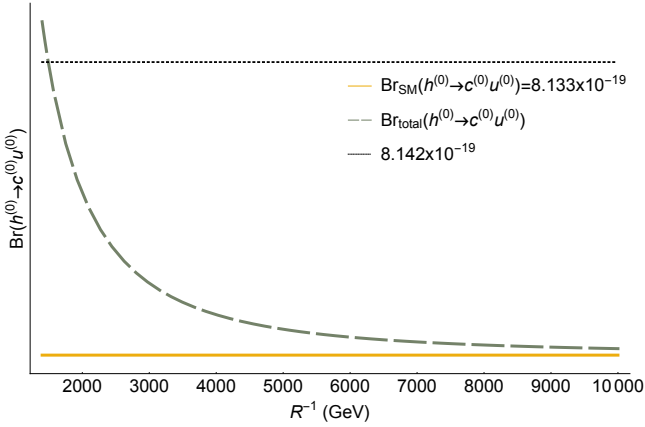}
\caption{\label{brcontrhcu2} Extra-dimensional effects on the branching ratio ${\rm Br}(h^{(0)}\to c^{(0)}u^{(0)})$. The graph includes a plot for the sole 4DSM contribution (lower horizontal line), and a plot for the total contribution from the whole KK theory (dashed curve).}
\end{figure}

Consider the Higgs decay $h^{(0)}\to b^{(0)}s^{(0)}$. The contributions to ${\rm Br}(h^{(0)}\to b^{(0)}s^{(0)})$, resulting from the full KK theory, have been plotted in Figs.~\ref{brcontrhbs1}-\ref{brcontrhbs2}. Fig.~\ref{brcontrhbs1} displays three plots,
with the upper solid curve representing the 4DSM prediction, the long-dashed plot corresponding to the interference of 4DSM and KK-excited-mode effects, and the lower short-dashed curve being associated to the contribution exclusively produced by KK excited modes. This graph has been plotted, in logarithmic scale, within the energy range that runs from $1.4\,{\rm TeV}$ to $5\,{\rm TeV}$. Dominance of the interference term over the pure KK-excited-modes contribution can be appreciated in this figure, which is expected since, according to Eqs.~(\ref{drateSMKK})-(\ref{serieshjKK}), leading interference effects on the branching ratio are of order $R^2$, whereas those from KK excited modes only are at least proportional to $R^4$. In this energy range, the difference between the 4DSM contribution and the leading extra-dimensional effects is about 2 orders of magnitude. For compactification-scale values $1.4\,{\rm TeV}<R^{-1}<10\,{\rm TeV}$, Fig.~\ref{brcontrhbs2} provides a comparison of the 4DSM branching ratio ${\rm Br}(h^{(0)}\to b^{(0)}s^{(0)})_{\rm SM}$ against the total branching ratio ${\rm Br}(h^{(0)}\to b^{(0)}s^{(0)})$, which is the sum of all KK contributions.
The solid horizontal lower line is the 4DSM contribution, whereas the dashed curve depicts the total branching ratio. For reference, an upper dotted horizontal line has been included, which corresponds to the value $1.82\times10^{-7}$. In accordance with the explicitly-decoupling behavior of effects from  KK excited modes, displayed by Eq.~(\ref{serieshjKK}), this graph shows how the total contribution gets increasingly similar to the 4DSM branching ratio as larger compactification scales are assumed. Regarding the quality of the estimations achieved through the $R$ series, valid for a large compactification scale, we provide Fig.~\ref{servslt},
where these results are compared to estimations obtained through direct numerical evaluation by usage of the package LoopTools. In this figure, only the interference of 4DSM and KK-excited-modes contributions has been considered, with the solid plot representing the LoopTools estimation and the dashed curve corresponding to the $R$-series branching-ratio term produced by Eqs.~(\ref{drateSMKK}) and (\ref{serieshjKK}). The LoopTools-estimation plot has been realized by taking only the contributions from the first three KK terms in the KK sum. The curves in Fig.~\ref{servslt}, plotted along the energy-interval $1.4\,{\rm TeV}<R^{-1}<5\,{\rm TeV}$, are in reasonable agreement with each other, thus illustrating that the $R$-series expression provides a good approximation. Analogous explanations to those given for Figs.~\ref{brcontrhbs1}-\ref{brcontrhbs2} apply for the graphs of Figs.~\ref{brcontrhbd1}-\ref{brcontrhcu2},
which provide characterizations of the KK contributions to ${\rm Br}(h^{(0)}\to b^{(0)}d^{(0)})$ and ${\rm Br}(h^{(0)}\to c^{(0)}u^{(0)})$. 
\\

\begin{figure}[ht]
\center
\includegraphics[width=8.5cm]{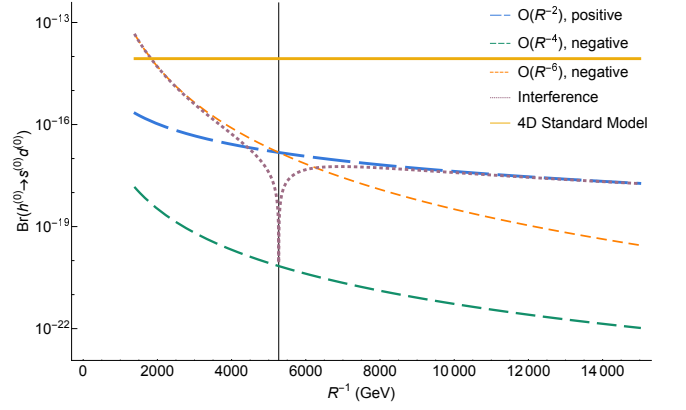}
\caption{\label{brcontrhcsd1} Comparison of $R^2$, $R^4$, and $R^6$ terms of ${\rm Br}(h^{(0)}\to s^{(0)}d^{(0)})$ within $1.4\,{\rm TeV}<R^{-1}<15\,{\rm TeV}$, plotted in logarithmic scale. The graph comprises plots for the 4DSM contribution (solid horizontal line), the full interference effect (dotted curve), the $R^2$ contribution (long-dashed curve), the $R^4$ contribution (medium-dashed curve), and the $R^6$ contribution (short-dashed curve).}
\end{figure}

\begin{figure}[ht]
\includegraphics[width=8.5cm]{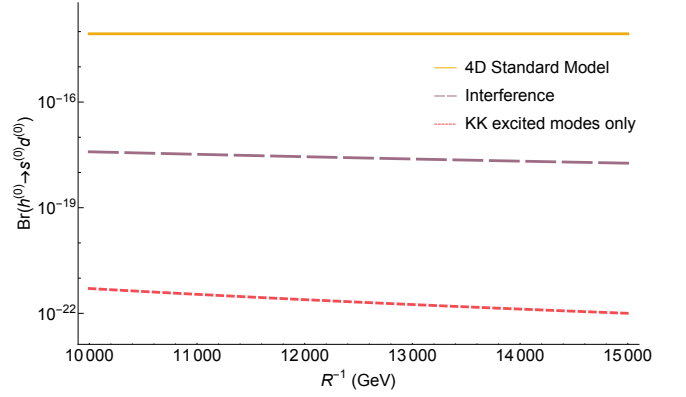}
\caption{\label{brcontrhcsd2} Terms of the branching ratio of $h^{(0)}\to s^{(0)}d^{(0)}$ within $10\,{\rm TeV}<R^{-1}<15\,{\rm TeV}$, plotted in logarithmic scale. Such terms are the 4DSM contribution (solid horizontal line), the interference of 4DSM and extra-dimensional contributions (large-dashed curve), and the contribution from KK excited modes only (short-dashed curve).}
\end{figure}

Regarding the branching ratio for $h^{(0)}\to s^{(0)}d^{(0)}$, an energy range of larger compactification scales should be considered for the $R$-series expression given in Eq.~(\ref{serieshjKK}) to yield a reliable prediction. To illustrate this, consider Figs.~\ref{brcontrhcsd1}-\ref{brcontrhcsd2}, which comprise two graphs of branching ratio as a function of the compactification scale $R^{-1}$, both plotted in logarithmic scale. The graph of Fig.~\ref{brcontrhcsd1},
carried out over the energy interval $1.4\,{\rm TeV}<R^{-1}<15\,{\rm TeV}$, includes a variety of plots, among which the upper horizontal solid line represents the 4DSM branching ratio, whereas the dotted curve represents the absolute value of the interference term, which we discuss because it is dominant over the pure-KK-excited-mode contribution. The interference term is a power series in $R^2$, from which we have identified the terms of orders $R^2$, $R^4$, and $R^6$, and plotted their absolute values in Fig.~\ref{brcontrhcsd1}. The long-dashed curve is associated to $R^2$ terms, whereas the medium-dashed and the short-dashed plots represent the contributions from the $R^4$ and $R^6$ terms, respectively. From this graph, notice that the $R^6$ term, which is negative, is the leading contribution for $R^{-1}\lesssim5.27\,{\rm TeV}$, even being dominant over the $R^2$ term, which is positive. A vertical solid line has been added to this graph with the objective of characterizing a change of sign of the interference contribution, from negative to positive, as for larger compactification scales the $R^2$ contribution becomes leading. This behavior suggests that the $R$ series is not convergent for $R^{-1}\lesssim5.27\,{\rm TeV}$. Taking this graph as a reference, we notice that from $R^{-1}\sim10\,{\rm TeV}$ and on the leading contribution is practically determined by the $R^2$ term. Thus we plotted Fig.~\ref{brcontrhcsd2},
which, for $10\,{\rm TeV}<R^{-1}<15\,{\rm TeV}$, provides a comparison among the contributions to the branching ratio ${\rm Br}(h^{(0)}\to s^{(0)}d^{(0)})$ given by the 4DSM contribution (solid horizontal line), the interference term (long-dashed curve), and the pure-KK-excited-modes contribution (short-dashed curve). 
\\


\section{Conclusions}
\label{conclusions}
The present investigation addressed the calculation and estimation of the effects on quark-flavor-changing Higgs-boson decays $h^{(0)}\to q^{(0)}_\alpha q^{(0)}_\beta$ generated by the Standard Model in five spacetime dimensions, with the assumption that the extra dimension is spacelike and universal. This hypothetical extra dimension is integrated out from the action of the five-dimensional Standard Model, then generating a four-dimensional effective Lagrangian that extends the four-dimensional Standard Model by the presence of Kaluza-Klein excited modes and which introduces the compactification scale, $R^{-1}$, as a new parameter. Moreover, the transition from five to four spacetime dimensions splits five-dimensional gauge invariance into two independent symmetries of the Kaluza-Klein theory, which are characterized the standard and the nonstandard gauge transformations. The calculation of ${\rm Br}(h^{(0)}\to q^{(0)}_\alpha q^{(0)}_\beta)$ involved a set of gauge-dependent Feynman diagrams which, contributing to $S$-matrix elements, add together to produce gauge-independent contributions. In this context, a ${\rm SU}(2,{\cal M}^4)_L\times{\rm U}(1,{\cal M}^4)_Y$-covariant nonlinear gauge was implemented to fix the gauge with respect to the nonstandard gauge transformations, whereas the unitary gauge was chosen to remove gauge invariance under standard gauge transformations. The new-physics effects were found to be UV finite and decoupling in the limit as $R^{-1}\to\infty$. A scenario characterized by a very small compactfication radius was then considered, which allowed for the derivation of brief analytic expressions for the form factors defining the branching-ratio formula for the decays under consideration. The usefulness of these results was discussed and numerical estimations were derived. We found that Kaluza-Klein leading contributions to both ${\rm Br}(h^{(0)}\to b^{(0)}s^{(0)})$ and ${\rm Br}(h^{(0)}\to b^{(0)}d^{(0)})$ are about 2 orders of magnitude below the four-dimensional-Standard-Model prediction if the compactification scale ranges within $1.4\,{\rm TeV}<R^{-1}<5\,{\rm TeV}$, whereas dominant extra-dimensional effects are at least 3 orders of magnitude smaller than those from the four-dimensional Standard Model if the energy interval $10\,{\rm TeV}<R^{-1}<1.5\,{\rm TeV}$ is considered. Regarding ${\rm Br}(h^{(0)}\to c^{(0)}u^{(0)})$, Kaluza-Klein contributions turned out to be smaller than those from the Standard Model in 4 dimensions by about 3 orders of magnitude as long as $1.4\,{\rm TeV}<R^{-1}<5\,{\rm TeV}$.




\begin{acknowledgments}
The authors acknowledge financial support from CONACYT and SNI (M\'exico).
\end{acknowledgments}


\appendix

\section{Coefficients of the amplitude $\Gamma^{\alpha\beta}$}
\label{App}
The amplitude $\Gamma^{\alpha\beta}$, given in Eq.~(\ref{Mamplitude}), is defined in terms of factors $h^{(k)\gamma}_{j,\alpha\beta}$, which are written in terms of two-point and three-point Passarino-Veltman scalar functions~\cite{PassVe,HooVe}, usually denoted as $B_0$ and $C_0$, respectively. In this Appendix, the expressions of these factors are displayed explicitly. Such expressions correspond to the case in which final-state quarks are zero-mode $u$-type quarks. The results for final-state $d$-type quarks are obtained from the equations provided in this Appendix by replacing $m_{u^{(0)}_\alpha}\to m_{d^{(0)}_\alpha}$, $m_{u^{(0)}_\beta}\to m_{d^{(0)}_\beta}$, and $m_{d^{(k)}_\gamma}\to m_{u^{(0)}_\gamma}$. Defining,
\begin{equation}
\zeta^{(k)}_{\alpha\beta}=\frac{ig^3}{256\,\pi^2\,m_{u^{(0)}_\alpha}m_{u^{(0)}_\beta}(m^2_{u^{(0)}_\alpha}-m^2_{u^{(0)}_\beta})m^3_{W^{(0)}}m^4_{W^{(k)}}},
\end{equation}
we write
\begin{widetext}
\begin{eqnarray}
&&
h^{(k)\gamma}_{1,\alpha\beta}=\frac{\zeta^{(k)}_{\alpha\beta}}{(m_{u^{(0)}_\alpha}+m_{u^{(0)}_\beta})^2-m^2_{h^{(0)}}}
\bigg[
-2B_0(0,m^2_{W^{(k)}},m^2_{W^{(k)}})m_{u^{(0)}_\alpha}(m_{u^{(0)}_\alpha}-m_{u^{(0)}_\beta})m_{u^{(0)}_\beta}
\nonumber \\ && \times
\big(-m^2_{h^{(0)}}+(m_{u^{(0)}_\alpha}+m_{u^{(0)}_\beta})^2\big)\Big(m_{W^{(k)}}^4+2(m_{W^{(0)}}^2-m_{(k)}^2)m_{W^{(k)}}^2+(m_{d^{(k)}_\gamma}^2+m_{u^{(0)}_\alpha} m_{u^{(0)}_\beta})
\nonumber \\ &&\times
(m_{(k)}^2+m_{W^{(0)}}^2)\Big)
   m_{W^{(k)}}^4+2 B_0(0,m^2_{d^{(k)}_\gamma},m^2_{d^{(k)}_\gamma})m_{d^{(k)}_\gamma}^2 m_{u^{(0)}_\alpha}(m_{u^{(0)}_\alpha}-m_{u^{(0)}_\beta})m_{u^{(0)}_\beta}\big(-m^2_{h^{(0)}}
\nonumber \\ &&
+(m_{u^{(0)}_\alpha}+m_{u^{(0)}_\beta})^2\big)\Big(m_{W^{(k)}}^4+2(m_{W^{(0)}}^2-m_{(k)}^2)m_{W^{(k)}}^2+(m_{d^{(k)}_\gamma}^2+m_{u^{(0)}_\alpha} m_{u^{(0)}_\beta})(m_{(k)}^2
\nonumber \\ &&
+m_{W^{(0)}}^2)\Big)m_{W^{(k)}}^2+B_0(0,m^2_{d^{(k)}_\gamma},m^2_{W^{(k)}})(m_{u^{(0)}_\alpha}-m_{u^{(0)}_\beta})\big(-m^2_{h^{(0)}}+(m_{u^{(0)}_\alpha}+m_{u^{(0)}_\beta})^2\big)
\nonumber \\ && \times
\big(m_{u^{(0)}_\alpha}^2+m_{u^{(0)}_\beta}^2\big)(m_{d^{(k)}_\gamma}-m_{W^{(k)}})(m_{d^{(k)}_\gamma}+m_{W^{(k)}})\Big(m_{W^{(k)}}^4+2(m_{W^{(0)}}^2-m_{(k)}^2)m_{W^{(k)}}^2
\nonumber \\ &&
+(m_{d^{(k)}_\gamma}^2+m_{u^{(0)}_\alpha} m_{u^{(0)}_\beta})(m_{(k)}^2+m_{W^{(0)}}^2)\Big)m_{W^{(k)}}^2+2 m_{u^{(0)}_\alpha}(m_{u^{(0)}_\alpha}-m_{u^{(0)}_\beta})m_{u^{(0)}_\beta} \big((m_{u^{(0)}_\alpha}+m_{u^{(0)}_\beta})^2
\nonumber \\ &&
-m_{h^{(0)}}^2\big)(m_{d^{(k)}_\gamma}-m_{W^{(k)}})(m_{d^{(k)}_\gamma}+m_{W^{(k)}})\Big(m_{W^{(k)}}^4+2(m_{W^{(0)}}^2-m_{(k)}^2)m_{W^{(k)}}^2+(m_{d^{(k)}_\gamma}^2
\nonumber \\ &&
+m_{u^{(0)}_\alpha} m_{u^{(0)}_\beta})(m_{(k)}^2+m_{W^{(0)}}^2)\Big)m_{W^{(k)}}^2-2 B_0(m^2_{h^{(0)}},m^2_{d^{(k)}_\gamma},m^2_{d^{(k)}_\gamma})m_{d^{(0)}_\gamma}^2 m_{u^{(0)}_\alpha}(m_{u^{(0)}_\alpha}-m_{u^{(0)}_\beta})m_{u^{(0)}_\beta}
\nonumber \\ && \times
(m_{u^{(0)}_\alpha}+m_{u^{(0)}_\beta})^2 \Big(2m_{W^{(k)}}^4-4 (m_{(k)}-m_{W^{(0)}})(m_{(k)}+m_{W^{(0)}}) m_{W^{(k)}}^2+(2m_{d^{(k)}_\gamma}^2-m_{h^{(0)}}^2
\nonumber \\ &&
+2m_{u^{(0)}_\alpha} m_{u^{(0)}_\beta})(m_{(k)}^2+m_{W^{(0)}}^2)\Big)m_{W^{(k)}}^2+2 C_0(m^2_{h^{(0)}},m^2_{u^{(0)}_\alpha},m^2_{u^{(0)}_\beta},m^2_{d^{(k)}_\gamma},m^2_{d^{(k)}_\gamma},m^2_{W^{(k)}})m_{d^{(0)}_\gamma}^2 m_{u^{(0)}_\alpha}
\nonumber \\ && \times
(m_{u^{(0)}_\alpha}-m_{u^{(0)}_\beta})m_{u^{(0)}_\beta} (m_{u^{(0)}_\alpha}+m_{u^{(0)}_\beta})^2 \Big(-2m_{W^{(k)}}^6+(2 m_{d^{(k)}_\gamma}^2-m_{h^{(0)}}^2+4m_{(k)}^2+m_{u^{(0)}_\alpha}^2+m_{u^{(0)}_\beta}^2
\nonumber \\ &&
-4m_{W^{(0)}}^2)m_{W^{(k)}}^4+\big((-6 m_{d^{(k)}_\gamma}^2+m_{h^{(0)}}^2+2 m_{u^{(0)}_\alpha}m_{u^{(0)}_\beta}) m_{(k)}^2+\big(2m_{d^{(k)}_\gamma}^2-m_{h^{(0)}}^2+2(m_{u^{(0)}_\alpha}^2
\nonumber \\ &&
-m_{u^{(0)}_\beta}m_{u^{(0)}_\alpha}+m_{u^{(0)}_\beta}^2)\big) m_{W^{(0)}}^2\big)m_{W^{(k)}}^2+(m_{d^{(k)}_\gamma}^2-m_{u^{(0)}_\alpha} m_{u^{(0)}_\beta})(2 m_{d^{(k)}_\gamma}^2-m_{u^{(0)}_\alpha}^2-m_{u^{(0)}_\beta}^2)(m_{(k)}^2
\nonumber \\ &&
+m_{W^{(0)}}^2)\Big)m_{W^{(k)}}^2+2 B_0(m^2_{h^{(0)}},m^2_{W^{(k)}},m^2_{W^{(k)}})m_{u^{(0)}_\alpha} (m_{u^{(0)}_\alpha}-m_{u^{(0)}_\beta}) m_{u^{(0)}_\beta}(m_{u^{(0)}_\alpha}+m_{u^{(0)}_\beta})^2
\nonumber \\ &&\times
\Big(2 m_{W^{(0)}}^2 m_{W^{(k)}}^6+(4m_{W^{(0)}}^4-4 m_{(k)}^2m_{W^{(0)}}^2+m_{h^{(0)}}^2 m_{(k)}^2)m_{W^{(k)}}^4-2(m_{(k)}^2+m_{W^{(0)}}^2)\big(m_{h^{(0)}}^2(m_{(k)}^2
\nonumber \\ &&
+m_{W^{(0)}}^2)-(m_{d^{(k)}_\gamma}^2+m_{u^{(0)}_\alpha}m_{u^{(0)}_\beta}) m_{W^{(0)}}^2\big)m_{W^{(k)}}^2+m_{h^{(0)}}^2(m_{d^{(k)}_\gamma}^2+m_{u^{(0)}_\alpha} m_{u^{(0)}_\beta})(m_{(k)}^2+m_{W^{(0)}}^2)^2\Big)
\nonumber \\ &&
+2C_0(m^2_{h^{(0)}},m^2_{u^{(0)}_\alpha},m^2_{u^{(0)}_\beta},m^2_{W^{(k)}},m^2_{W^{(k)}},m^2_{d^{(k)}_\gamma}) m_{u^{(0)}_\alpha}(m_{u^{(0)}_\alpha}-m_{u^{(0)}_\beta})m_{u^{(0)}_\beta} (m_{u^{(0)}_\alpha}+m_{u^{(0)}_\beta})^2
\nonumber \\ && \times
\Big(-2 m_{W^{(0)}}^2m_{W^{(k)}}^8+\big(2 m_{W^{(0)}}^2 (2m_{(k)}^2-2 m_{W^{(0)}}^2+m_{u^{(0)}_\alpha}m_{u^{(0)}_\beta})-m_{h^{(0)}}^2m_{(k)}^2\big) m_{W^{(k)}}^6+\big(2(m_{h^{(0)}}^2
\nonumber \\ &&
+m_{u^{(0)}_\alpha}m_{u^{(0)}_\beta})m_{W^{(0)}}^4+2m_{(k)}^2(m_{h^{(0)}}^2+m_{u^{(0)}_\alpha}^2+m_{u^{(0)}_\beta}^2-m_{u^{(0)}_\alpha}m_{u^{(0)}_\beta})m_{W^{(0)}}^2+m_{h^{(0)}}^2 m_{(k)}^2(2m_{(k)}^2
\nonumber \\ &&
+m_{u^{(0)}_\alpha}m_{u^{(0)}_\beta})\big)m_{W^{(k)}}^4-(m_{(k)}^2+m_{W^{(0)}}^2)\big(m_{h^{(0)}}^2(m_{h^{(0)}}^2-m_{u^{(0)}_\alpha}^2-m_{u^{(0)}_\beta}^2+m_{u^{(0)}_\alpha}m_{u^{(0)}_\beta})m_{(k)}^2
\nonumber \\ &&
+(m_{h^{(0)}}^2 (2m_{u^{(0)}_\alpha}^2-m_{u^{(0)}_\beta}m_{u^{(0)}_\alpha}+2 m_{u^{(0)}_\beta}^2)-2(m_{u^{(0)}_\alpha}^4-m_{u^{(0)}_\beta}^2 m_{u^{(0)}_\alpha}^2+m_{u^{(0)}_\beta}^4))m_{W^{(0)}}^2\big) m_{W^{(k)}}^2
\nonumber \\ &&
+m_{h^{(0)}}^2m_{u^{(0)}_\alpha}^2 m_{u^{(0)}_\beta}^2(m_{(k)}^2+m_{W^{(0)}}^2)^2+m_{d^{(k)}_\gamma}^4(m_{(k)}^2+m_{W^{(0)}}^2)\big((m_{(k)}^2+m_{W^{(0)}}^2)m_{h^{(0)}}^2
\nonumber \\ &&
+2 m_{W^{(0)}}^2m_{W^{(k)}}^2\big)+m_{d^{(k)}_\gamma}^2\big(2 m_{W^{(0)}}^2 m_{W^{(k)}}^6+(2m_{W^{(0)}}^4-6 m_{(k)}^2m_{W^{(0)}}^2+m_{h^{(0)}}^2 m_{(k)}^2)m_{W^{(k)}}^4-(m_{(k)}^2
\nonumber \\ &&
+m_{W^{(0)}}^2) ((3m_{(k)}^2+m_{W^{(0)}}^2)m_{h^{(0)}}^2+2(m_{u^{(0)}_\alpha}^2+m_{u^{(0)}_\beta}^2) m_{W^{(0)}}^2)m_{W^{(k)}}^2+m_{h^{(0)}}^2(m_{h^{(0)}}^2-m_{u^{(0)}_\alpha}^2
\nonumber \\ &&
-m_{u^{(0)}_\beta}^2)(m_{(k)}^2+m_{W^{(0)}}^2)^2\big)\Big)+B_0(m^2_{u^{(0)}_\alpha},m^2_{d^{(k)}_\gamma},m^2_{W^{(k)}}) m_{u^{(0)}_\beta}(m_{u^{(0)}_\alpha}+m_{u^{(0)}_\beta})\Big(m_{u^{(0)}_\beta}(-m^2_{h^{(0)}}
\nonumber \\ &&
+(m_{u^{(0)}_\alpha}+m_{u^{(0)}_\beta})^2)(m_{(k)}^2+m_{W^{(0)}}^2)m_{W^{(k)}}^2 m_{d^{(k)}_\gamma}^4+\big(m_{u^{(0)}_\beta}(-m^2_{h^{(0)}}+(m_{u^{(0)}_\alpha}+m_{u^{(0)}_\beta})^2)m_{W^{(k)}}^6
\nonumber \\ &&
+(m_{h^{(0)}}-m_{u^{(0)}_\alpha}-m_{u^{(0)}_\beta}) m_{u^{(0)}_\beta}(m_{h^{(0)}}+m_{u^{(0)}_\alpha}+m_{u^{(0)}_\beta}) (3m_{(k)}^2-m_{W^{(0)}}^2)m_{W^{(k)}}^4-m_{u^{(0)}_\alpha}(m_{(k)}^2
\nonumber \\ &&
+m_{W^{(0)}}^2)(4m_{u^{(0)}_\alpha} (m_{u^{(0)}_\beta}-m_{u^{(0)}_\alpha}) m_{d^{(0)}_\gamma}^2+4 m_{u^{(0)}_\alpha} (m_{u^{(0)}_\alpha}-m_{u^{(0)}_\beta})m_{W^{(0)}}^2+m_{u^{(0)}_\beta}(m_{u^{(0)}_\alpha}+m_{u^{(0)}_\beta})
\nonumber \\  && \times
(-m^2_{h^{(0)}}+(m_{u^{(0)}_\alpha}+m_{u^{(0)}_\beta})^2)) m_{W^{(k)}}^2-2m_{h^{(0)}}^2 m_{u^{(0)}_\alpha}^2(m_{u^{(0)}_\alpha}-m_{u^{(0)}_\beta})(m_{(k)}^2+m_{W^{(0)}}^2)^2\big)m_{d^{(k)}_\gamma}^2
\nonumber \\ &&
+m_{h^{(0)}}^2\big(m_{u^{(0)}_\beta}m_{W^{(k)}}^8-m_{u^{(0)}_\beta} (2m_{(k)}^2+m_{u^{(0)}_\alpha}^2-2m_{W^{(0)}}^2)m_{W^{(k)}}^6+m_{u^{(0)}_\alpha}(m_{u^{(0)}_\beta} (m_{u^{(0)}_\beta}
\nonumber \\ &&
-2m_{u^{(0)}_\alpha}) m_{W^{(0)}}^2-m_{(k)}^2(2 m_{u^{(0)}_\alpha}^2-2m_{u^{(0)}_\beta} m_{u^{(0)}_\alpha}+m_{u^{(0)}_\beta}^2))m_{W^{(k)}}^4+m_{u^{(0)}_\alpha}(m_{(k)}^2+m_{W^{(0)}}^2)(4m_{(k)}^2
\nonumber \\ &&\times
m_{u^{(0)}_\alpha}(m_{u^{(0)}_\alpha}-m_{u^{(0)}_\beta})-m_{u^{(0)}_\beta}(m_{u^{(0)}_\beta} m_{u^{(0)}_\alpha}^2+4 (m_{u^{(0)}_\beta}-m_{u^{(0)}_\alpha})m_{W^{(0)}}^2)) m_{W^{(k)}}^2-2m_{u^{(0)}_\alpha}^3 (m_{u^{(0)}_\alpha}
\nonumber \\ &&
-m_{u^{(0)}_\beta}) m_{u^{(0)}_\beta}(m_{(k)}^2+m_{W^{(0)}}^2)^2\big)+m_{W^{(k)}}^2\big(-m_{u^{(0)}_\beta}(m_{u^{(0)}_\alpha}+m_{u^{(0)}_\beta})^2m_{W^{(k)}}^6+((2m_{(k)}^2+m_{u^{(0)}_\alpha}^2)
\nonumber \\ && \times
m_{u^{(0)}_\beta} (m_{u^{(0)}_\alpha}+m_{u^{(0)}_\beta})^2-2(2m_{u^{(0)}_\alpha}^3-m_{u^{(0)}_\beta}m_{u^{(0)}_\alpha}^2+2 m_{u^{(0)}_\beta}^2m_{u^{(0)}_\alpha}+m_{u^{(0)}_\beta}^3) m_{W^{(0)}}^2)m_{W^{(k)}}^4+m_{u^{(0)}_\alpha}
\nonumber \\ && \times
(8m_{u^{(0)}_\alpha} (m_{u^{(0)}_\beta}-m_{u^{(0)}_\alpha}) m_{W^{(0)}}^4+(2m_{u^{(0)}_\alpha}-m_{u^{(0)}_\beta})m_{u^{(0)}_\beta} (m_{u^{(0)}_\alpha}+m_{u^{(0)}_\beta})^2m_{W^{(0)}}^2+m_{(k)}^2
\nonumber \\ && \times
(m_{u^{(0)}_\beta}^2 (m_{u^{(0)}_\alpha}+m_{u^{(0)}_\beta})^2+8 m_{u^{(0)}_\alpha} (m_{u^{(0)}_\alpha}-m_{u^{(0)}_\beta})m_{W^{(0)}}^2))m_{W^{(k)}}^2+m_{u^{(0)}_\alpha}(m_{(k)}^2+m_{W^{(0)}}^2)(m_{u^{(0)}_\alpha}^2
\nonumber \\ && \times
m_{u^{(0)}_\beta}^2 (m_{u^{(0)}_\alpha}+m_{u^{(0)}_\beta})^2+4(m_{u^{(0)}_\alpha}^4-m_{u^{(0)}_\beta} m_{u^{(0)}_\alpha}^3-m_{u^{(0)}_\beta}^2 m_{u^{(0)}_\alpha}^2+m_{u^{(0)}_\beta}^4)m_{W^{(0)}}^2)-2 m_{d^{(0)}_\gamma}^2m_{u^{(0)}_\alpha}^2
\nonumber \\ && \times
(m_{u^{(0)}_\alpha}-m_{u^{(0)}_\beta})(-2 m_{W^{(k)}}^4+4(m_{(k)}-m_{W^{(0)}}) (m_{(k)}+m_{W^{(0)}})m_{W^{(k)}}^2+(m_{u^{(0)}_\alpha}^2+m_{u^{(0)}_\beta}^2)(m_{(k)}^2
\nonumber \\ &&
+m_{W^{(0)}}^2))\big)\Big)-B_0(m^2_{u^{(0)}_\beta},m^2_{d^{(k)}_\gamma},m^2_{W^{(k)}})m_{u^{(0)}_\alpha} (m_{u^{(0)}_\alpha}+m_{u^{(0)}_\beta})\Big(m_{u^{(0)}_\alpha} (-m^2_{h^{(0)}}+(m_{u^{(0)}_\alpha}+m_{u^{(0)}_\beta})^2)
\nonumber \\ && \times
(m_{(k)}^2+m_{W^{(0)}}^2)m_{W^{(k)}}^2 m_{d^{(k)}_\gamma}^4+\big(m_{u^{(0)}_\alpha}(-m_{h^{(0)}}+m_{u^{(0)}_\alpha}+m_{u^{(0)}_\beta})(m_{h^{(0)}}+m_{u^{(0)}_\alpha}+m_{u^{(0)}_\beta})m_{W^{(k)}}^6
\nonumber \\ &&
-m_{u^{(0)}_\alpha}(-m_{h^{(0)}}+m_{u^{(0)}_\alpha}+m_{u^{(0)}_\beta})(m_{h^{(0)}}+m_{u^{(0)}_\alpha}+m_{u^{(0)}_\beta}) \left(3m_{(k)}^2-m_{W^{(0)}}^2\right)m_{W^{(k)}}^4-m_{u^{(0)}_\beta}
\nonumber \\ && \times
(m_{(k)}^2+m_{W^{(0)}}^2)(m_{u^{(0)}_\alpha}^4+3 m_{u^{(0)}_\beta} m_{u^{(0)}_\alpha}^3+3 m_{u^{(0)}_\beta}^2 m_{u^{(0)}_\alpha}^2-m_{h^{(0)}}^2(m_{u^{(0)}_\alpha}+m_{u^{(0)}_\beta})m_{u^{(0)}_\alpha}+m_{u^{(0)}_\beta} (4m_{d^{(0)}_\gamma}^2
\nonumber \\ &&
+m_{u^{(0)}_\beta}^2-4m_{W^{(0)}}^2) m_{u^{(0)}_\alpha}+4m_{u^{(0)}_\beta}^2(m_{W^{(0)}}^2-m_{d^{(0)}_\gamma}^2)) m_{W^{(k)}}^2+2 m_{h^{(0)}}^2(m_{u^{(0)}_\alpha}-m_{u^{(0)}_\beta})m_{u^{(0)}_\beta}^2(m_{(k)}^2
\nonumber \\ &&
+m_{W^{(0)}}^2)^2\big)m_{d^{(k)}_\gamma}^2+m_{h^{(0)}}^2\big(m_{u^{(0)}_\alpha}m_{W^{(k)}}^8-m_{u^{(0)}_\alpha} (2m_{(k)}^2+m_{u^{(0)}_\beta}^2-2m_{W^{(0)}}^2)m_{W^{(k)}}^6+m_{u^{(0)}_\beta}\big(m_{u^{(0)}_\alpha}
\nonumber \\ && \times
(m_{u^{(0)}_\alpha}-2 m_{u^{(0)}_\beta})m_{W^{(0)}}^2-m_{(k)}^2(m_{u^{(0)}_\alpha}^2-2 m_{u^{(0)}_\beta} m_{u^{(0)}_\alpha}+2 m_{u^{(0)}_\beta}^2)\big)m_{W^{(k)}}^4+m_{u^{(0)}_\beta}(m_{(k)}^2+m_{W^{(0)}}^2)
\nonumber \\ && \times
(4m_{(k)}^2 m_{u^{(0)}_\beta}(m_{u^{(0)}_\beta}-m_{u^{(0)}_\alpha})-m_{u^{(0)}_\alpha}(m_{u^{(0)}_\alpha} m_{u^{(0)}_\beta}^2+4 (m_{u^{(0)}_\alpha}-m_{u^{(0)}_\beta})m_{W^{(0)}}^2)) m_{W^{(k)}}^2+2m_{u^{(0)}_\alpha}
\nonumber \\ && \times
(m_{u^{(0)}_\alpha}-m_{u^{(0)}_\beta}) m_{u^{(0)}_\beta}^3(m_{(k)}^2+m_{W^{(0)}}^2)^2\big)+m_{W^{(k)}}^2 \big(-m_{u^{(0)}_\alpha}(m_{u^{(0)}_\alpha}+m_{u^{(0)}_\beta})^2m_{W^{(k)}}^6+\big(m_{u^{(0)}_\alpha}
\nonumber \\ && \times
(m_{u^{(0)}_\alpha}+m_{u^{(0)}_\beta})^2(2 m_{(k)}^2+m_{u^{(0)}_\beta}^2)-2 (m_{u^{(0)}_\alpha}^3+2m_{u^{(0)}_\beta} m_{u^{(0)}_\alpha}^2-m_{u^{(0)}_\beta}^2 m_{u^{(0)}_\alpha}+2 m_{u^{(0)}_\beta}^3)m_{W^{(0)}}^2\big)m_{W^{(k)}}^4
\nonumber \\ &&
+m_{u^{(0)}_\beta}(8(m_{u^{(0)}_\alpha}-m_{u^{(0)}_\beta})m_{u^{(0)}_\beta}m_{W^{(0)}}^4-m_{u^{(0)}_\alpha}(m_{u^{(0)}_\alpha}-2 m_{u^{(0)}_\beta})(m_{u^{(0)}_\alpha}+m_{u^{(0)}_\beta})^2m_{W^{(0)}}^2+m_{(k)}^2
\nonumber \\ && \times
(m_{u^{(0)}_\alpha}^2 (m_{u^{(0)}_\alpha}+m_{u^{(0)}_\beta})^2+8 m_{u^{(0)}_\beta}(m_{u^{(0)}_\beta}-m_{u^{(0)}_\alpha})m_{W^{(0)}}^2))m_{W^{(k)}}^2+m_{u^{(0)}_\beta}(m_{(k)}^2+m_{W^{(0)}}^2)(m_{u^{(0)}_\alpha}^2
\nonumber \\ && \times
m_{u^{(0)}_\beta}^2 (m_{u^{(0)}_\alpha}+m_{u^{(0)}_\beta})^2+4 (m_{u^{(0)}_\alpha}-m_{u^{(0)}_\beta})(m_{u^{(0)}_\alpha}^3+m_{u^{(0)}_\beta} m_{u^{(0)}_\alpha}^2-m_{u^{(0)}_\beta}^3)m_{W^{(0)}}^2)+2 m_{d^{(0)}_\gamma}^2(m_{u^{(0)}_\alpha}
\nonumber \\ &&
-m_{u^{(0)}_\beta})m_{u^{(0)}_\beta}^2(-2 m_{W^{(k)}}^4+4(m_{(k)}-m_{W^{(0)}}) (m_{(k)}+m_{W^{(0)}})m_{W^{(k)}}^2
\nonumber \\ &&
+(m_{u^{(0)}_\alpha}^2+m_{u^{(0)}_\beta}^2)(m_{(k)}^2+m_{W^{(0)}}^2))\big)\Big)
\bigg],
\end{eqnarray}
\begin{eqnarray}
&&
h^{(k)\gamma}_{2,\alpha\beta}=\frac{\zeta^{(k)}_{\alpha\beta}}{(m_{u^{(0)}_\alpha}-m_{u^{(0)}_\beta})^2-m^2_{h^{(0)}}}
\bigg[
-2B_0(0,m^2_{W^{(k)}},m^2_{W^{(k)}})m_{u^{(0)}_\alpha}(-m^2_{h^{(0)}}+(m_{u^{(0)}_\alpha}-m_{u^{(0)}_\beta})^2)m_{u^{(0)}_\beta}
\nonumber \\ && \times
(m_{u^{(0)}_\alpha}+m_{u^{(0)}_\beta})\big(-m_{W^{(k)}}^4+2 (m^2_{(k)}-m^2_{W^{(0)}})m_{W^{(k)}}^2-(m_{d^{(k)}_\gamma}^2-m_{u^{(0)}_\alpha} m_{u^{(0)}_\beta})(m_{(k)}^2
+m_{W^{(0)}}^2)\big)m_{W^{(k)}}^4
\nonumber \\ &&
-2 B_0(m^2_{h^{(0)}},m^2_{d^{(k)}_\gamma},m^2_{d^{(k)}_\gamma})m_{d^{(0)}_\gamma}^2 m_{u^{(0)}_\alpha}(m_{u^{(0)}_\alpha}-m_{u^{(0)}_\beta})^2m_{u^{(0)}_\beta} (m_{u^{(0)}_\alpha}+m_{u^{(0)}_\beta})
\big(-2 m_{W^{(k)}}^4+4(m^2_{(k)}
\nonumber \\ &&
-m^2_{W^{(0)}})m_{W^{(k)}}^2-(2 m_{d^{(k)}_\gamma}^2-m_{h^{(0)}}^2-2 m_{u^{(0)}_\alpha}m_{u^{(0)}_\beta})(m_{(k)}^2+m_{W^{(0)}}^2)\big)m_{W^{(k)}}^2-2 B_0(0,m^2_{d^{(k)}_\gamma},m^2_{d^{(k)}_\gamma})m_{d^{(k)}_\gamma}^2
\nonumber \\ && \times
m_{u^{(0)}_\alpha}(-m^2_{h^{(0)}}+(m_{u^{(0)}_\alpha}-m_{u^{(0)}_\beta})^2)m_{u^{(0)}_\beta}(m_{u^{(0)}_\alpha}+m_{u^{(0)}_\beta})\big(m_{W^{(k)}}^4-2 (m^2_{(k)}-m^2_{W^{(0)}})m_{W^{(k)}}^2+(m_{d^{(k)}_\gamma}^2
\nonumber \\ &&
-m_{u^{(0)}_\alpha} m_{u^{(0)}_\beta})(m_{(k)}^2+m_{W^{(0)}}^2)\big)m_{W^{(k)}}^2+2 m_{u^{(0)}_\alpha}(m_{h^{(0)}}+m_{u^{(0)}_\alpha}-m_{u^{(0)}_\beta}) m_{u^{(0)}_\beta}(m_{h^{(0)}}-m_{u^{(0)}_\alpha}+m_{u^{(0)}_\beta})
\nonumber \\ && \times
(m_{u^{(0)}_\alpha}+m_{u^{(0)}_\beta}) (m^2_{d^{(k)}_\gamma}-m^2_{W^{(k)}})\big(m_{W^{(k)}}^4-2(m^2_{(k)}-m^2_{W^{(0)}})m_{W^{(k)}}^2+(m_{d^{(k)}_\gamma}^2-m_{u^{(0)}_\alpha} m_{u^{(0)}_\beta})(m_{(k)}^2
\nonumber \\ &&
+m_{W^{(0)}}^2)\big)m_{W^{(k)}}^2-B_0(0,m^2_{d^{(k)}_\gamma},m^2_{W^{(k)}})(m_{h^{(0)}}+m_{u^{(0)}_\alpha}-m_{u^{(0)}_\beta})(m_{h^{(0)}}-m_{u^{(0)}_\alpha}+m_{u^{(0)}_\beta}) (m_{u^{(0)}_\alpha}+m_{u^{(0)}_\beta})
\nonumber \\ && \times
(m_{u^{(0)}_\alpha}^2+m_{u^{(0)}_\beta}^2)(m_{d^{(k)}_\gamma}-m_{W^{(k)}})(m_{d^{(k)}_\gamma}+m_{W^{(k)}})\big(m_{W^{(k)}}^4-2 (m_{(k)}-m_{W^{(0)}})(m_{(k)}+m_{W^{(0)}})m_{W^{(k)}}^2
\nonumber \\ &&
+(m_{d^{(k)}_\gamma}^2-m_{u^{(0)}_\alpha} m_{u^{(0)}_\beta})(m_{(k)}^2+m_{W^{(0)}}^2)\big)m_{W^{(k)}}^2+2 C_0(m^2_{h^{(0)}},m^2_{u^{(0)}_\alpha},m^2_{u^{(0)}_\beta},m^2_{d^{(k)}_\gamma},m^2_{d^{(k)}_\gamma},m^2_{W^{(k)}})m_{d^{(0)}_\gamma}^2
\nonumber \\ && \times
m_{u^{(0)}_\alpha}(m_{u^{(0)}_\alpha}-m_{u^{(0)}_\beta})^2m_{u^{(0)}_\beta} (m_{u^{(0)}_\alpha}+m_{u^{(0)}_\beta}) \Big(2m_{W^{(k)}}^6-(2 m_{d^{(k)}_\gamma}^2-m_{h^{(0)}}^2+4m_{(k)}^2+m_{u^{(0)}_\alpha}^2+m_{u^{(0)}_\beta}^2
\nonumber \\ &&
-4m_{W^{(0)}}^2)m_{W^{(k)}}^4+\big((6m_{d^{(k)}_\gamma}^2-m_{h^{(0)}}^2+2m_{u^{(0)}_\alpha} m_{u^{(0)}_\beta})m_{(k)}^2+(-2 m_{d^{(k)}_\gamma}^2+m_{h^{(0)}}^2-2(m_{u^{(0)}_\alpha}^2+m_{u^{(0)}_\beta} m_{u^{(0)}_\alpha}
\nonumber \\ &&
+m_{u^{(0)}_\beta}^2))m_{W^{(0)}}^2\big)m_{W^{(k)}}^2-(m_{d^{(k)}_\gamma}^2+m_{u^{(0)}_\alpha} m_{u^{(0)}_\beta})(2 m_{d^{(k)}_\gamma}^2-m_{u^{(0)}_\alpha}^2-m_{u^{(0)}_\beta}^2)(m_{(k)}^2+m_{W^{(0)}}^2)\Big)m_{W^{(k)}}^2
\nonumber \\ &&
+2B_0(m^2_{h^{(0)}},m^2_{W^{(k)}},m^2_{W^{(k)}})m_{u^{(0)}_\alpha} (m_{u^{(0)}_\alpha}-m_{u^{(0)}_\beta})^2 m_{u^{(0)}_\beta}(m_{u^{(0)}_\alpha}+m_{u^{(0)}_\beta})\Big(-2 m_{W^{(0)}}^2 m_{W^{(k)}}^6-(4m_{W^{(0)}}^4
\nonumber \\ &&
-4 m_{(k)}^2m_{W^{(0)}}^2+m_{h^{(0)}}^2 m_{(k)}^2)m_{W^{(k)}}^4+2(m_{(k)}^2+m_{W^{(0)}}^2)\big(m_{h^{(0)}}^2m_{(k)}^2+(-m_{d^{(k)}_\gamma}^2+m_{h^{(0)}}^2+m_{u^{(0)}_\alpha}m_{u^{(0)}_\beta}) m_{W^{(0)}}^2\big)
\nonumber \\ && \times
m_{W^{(k)}}^2-m_{h^{(0)}}^2(m_{d^{(k)}_\gamma}^2-m_{u^{(0)}_\alpha} m_{u^{(0)}_\beta})(m_{(k)}^2+m_{W^{(0)}}^2)^2\Big)-2 C_0(m^2_{h^{(0)}},m^2_{u^{(0)}_\alpha},m^2_{u^{(0)}_\beta},m^2_{W^{(k)}},m^2_{W^{(k)}},m^2_{d^{(k)}_\gamma})
\nonumber \\ && \times
m_{u^{(0)}_\alpha}(m_{u^{(0)}_\alpha}-m_{u^{(0)}_\beta})^2m_{u^{(0)}_\beta} (m_{u^{(0)}_\alpha}+m_{u^{(0)}_\beta}) \Big(-2 m_{W^{(0)}}^2m_{W^{(k)}}^8-\big(m_{h^{(0)}}^2 m_{(k)}^2+2m_{W^{(0)}}^2(-2 m_{(k)}^2+2m_{W^{(0)}}^2
\nonumber \\ &&
+m_{u^{(0)}_\alpha}m_{u^{(0)}_\beta})\big)m_{W^{(k)}}^6+\big(2(m_{h^{(0)}}^2-m_{u^{(0)}_\alpha}m_{u^{(0)}_\beta}) m_{W^{(0)}}^4+2m_{(k)}^2(m_{h^{(0)}}^2+m_{u^{(0)}_\alpha}^2+m_{u^{(0)}_\beta}^2+m_{u^{(0)}_\alpha}m_{u^{(0)}_\beta})m_{W^{(0)}}^2
\nonumber \\ &&
+m_{h^{(0)}}^2 m_{(k)}^2(2m_{(k)}^2-m_{u^{(0)}_\alpha}m_{u^{(0)}_\beta})\big)m_{W^{(k)}}^4-(m_{(k)}^2+m_{W^{(0)}}^2)\big(m_{(k)}^2m_{h^{(0)}}^4+((2 m_{u^{(0)}_\alpha}^2+m_{u^{(0)}_\beta} m_{u^{(0)}_\alpha}+2m_{u^{(0)}_\beta}^2)
\nonumber \\ && \times
m_{W^{(0)}}^2-m_{(k)}^2(m_{u^{(0)}_\alpha}^2+m_{u^{(0)}_\beta}m_{u^{(0)}_\alpha}+m_{u^{(0)}_\beta}^2)) m_{h^{(0)}}^2-2(m_{u^{(0)}_\alpha}^4-m_{u^{(0)}_\beta}^2 m_{u^{(0)}_\alpha}^2+m_{u^{(0)}_\beta}^4)m_{W^{(0)}}^2\big)m_{W^{(k)}}^2
\nonumber \\ &&
+m_{h^{(0)}}^2 m_{u^{(0)}_\alpha}^2 m_{u^{(0)}_\beta}^2(m_{(k)}^2+m_{W^{(0)}}^2)^2+m_{d^{(k)}_\gamma}^4(m_{(k)}^2+m_{W^{(0)}}^2)((m_{(k)}^2+m_{W^{(0)}}^2)m_{h^{(0)}}^2+2 m_{W^{(0)}}^2m_{W^{(k)}}^2)
\nonumber \\ &&
+m_{d^{(k)}_\gamma}^2\big(2 m_{W^{(0)}}^2 m_{W^{(k)}}^6+(2m_{W^{(0)}}^4-6 m_{(k)}^2m_{W^{(0)}}^2+m_{h^{(0)}}^2 m_{(k)}^2)m_{W^{(k)}}^4-(m_{(k)}^2+m_{W^{(0)}}^2) ((3m_{(k)}^2
\nonumber \\ &&
+m_{W^{(0)}}^2) m_{h^{(0)}}^2+2(m_{u^{(0)}_\alpha}^2+m_{u^{(0)}_\beta}^2)m_{W^{(0)}}^2)m_{W^{(k)}}^2+m_{h^{(0)}}^2(m_{h^{(0)}}^2-m_{u^{(0)}_\alpha}^2-m_{u^{(0)}_\beta}^2)(m_{(k)}^2+m_{W^{(0)}}^2)^2\big)\Big)
\nonumber \\ &&
-B_0(m^2_{u^{(0)}_\alpha},m^2_{d^{(k)}_\gamma},m^2_{W^{(k)}}) (m_{u^{(0)}_\alpha}-m_{u^{(0)}_\beta}) m_{u^{(0)}_\beta}\Big((m_{h^{(0)}}+m_{u^{(0)}_\alpha}-m_{u^{(0)}_\beta}) m_{u^{(0)}_\beta}(m_{h^{(0)}}-m_{u^{(0)}_\alpha}
\nonumber \\ &&
+m_{u^{(0)}_\beta})(m_{(k)}^2+m_{W^{(0)}}^2)m_{W^{(k)}}^2 m_{d^{(k)}_\gamma}^4-\big(m_{u^{(0)}_\beta}(-m_{h^{(0)}}-m_{u^{(0)}_\alpha}+m_{u^{(0)}_\beta})(m_{h^{(0)}}-m_{u^{(0)}_\alpha}+m_{u^{(0)}_\beta})m_{W^{(k)}}^6
\nonumber \\ &&
+(m_{h^{(0)}}+m_{u^{(0)}_\alpha}-m_{u^{(0)}_\beta}) m_{u^{(0)}_\beta}(m_{h^{(0)}}-m_{u^{(0)}_\alpha}+m_{u^{(0)}_\beta}) (3m_{(k)}^2-m_{W^{(0)}}^2)m_{W^{(k)}}^4+m_{u^{(0)}_\alpha}(m_{(k)}^2+m_{W^{(0)}}^2)
\nonumber \\ && \times
(-m_{u^{(0)}_\beta} (m_{u^{(0)}_\alpha}-m_{u^{(0)}_\beta})^3+m_{h^{(0)}}^2m_{u^{(0)}_\beta} (m_{u^{(0)}_\alpha}-m_{u^{(0)}_\beta})+4 m_{u^{(0)}_\alpha}(m_{u^{(0)}_\alpha}+m_{u^{(0)}_\beta})m_{W^{(0)}}^2-4 m_{d^{(0)}_\gamma}^2m_{u^{(0)}_\alpha}
\nonumber \\ && \times
(m_{u^{(0)}_\alpha}+m_{u^{(0)}_\beta}))m_{W^{(k)}}^2+2m_{h^{(0)}}^2 m_{u^{(0)}_\alpha}^2(m_{u^{(0)}_\alpha}+m_{u^{(0)}_\beta})(m_{(k)}^2+m_{W^{(0)}}^2)^2\big)m_{d^{(k)}_\gamma}^2+m_{h^{(0)}}^2\big(-m_{u^{(0)}_\beta}m_{W^{(k)}}^8
\nonumber \\ &&
+m_{u^{(0)}_\beta} (2m_{(k)}^2+m_{u^{(0)}_\alpha}^2-2m_{W^{(0)}}^2)m_{W^{(k)}}^6+m_{u^{(0)}_\alpha}(m_{u^{(0)}_\beta} (2 m_{u^{(0)}_\alpha}+m_{u^{(0)}_\beta})m_{W^{(0)}}^2-m_{(k)}^2(2m_{u^{(0)}_\alpha}^2
\nonumber \\ &&
+2m_{u^{(0)}_\beta}m_{u^{(0)}_\alpha}+m_{u^{(0)}_\beta}^2))m_{W^{(k)}}^4+m_{u^{(0)}_\alpha}(m_{(k)}^2+m_{W^{(0)}}^2)(4m_{u^{(0)}_\alpha} (m_{u^{(0)}_\alpha}+m_{u^{(0)}_\beta})m_{(k)}^2+m_{u^{(0)}_\beta}(-m_{u^{(0)}_\beta} m_{u^{(0)}_\alpha}^2
\nonumber \\ &&
-4(m_{u^{(0)}_\alpha}+m_{u^{(0)}_\beta}) m_{W^{(0)}}^2)) m_{W^{(k)}}^2+2m_{u^{(0)}_\alpha}^3 m_{u^{(0)}_\beta}(m_{u^{(0)}_\alpha}+m_{u^{(0)}_\beta})(m_{(k)}^2+m_{W^{(0)}}^2)^2\big)+m_{W^{(k)}}^2 \big((m_{u^{(0)}_\alpha}
\nonumber \\ &&
-m_{u^{(0)}_\beta})^2 m_{u^{(0)}_\beta}m_{W^{(k)}}^6-((2m_{(k)}^2+m_{u^{(0)}_\alpha}^2)m_{u^{(0)}_\beta} (m_{u^{(0)}_\alpha}-m_{u^{(0)}_\beta})^2+2(2m_{u^{(0)}_\alpha}^3+m_{u^{(0)}_\beta}m_{u^{(0)}_\alpha}^2+2 m_{u^{(0)}_\beta}^2m_{u^{(0)}_\alpha}
\nonumber \\ &&
-m_{u^{(0)}_\beta}^3)m_{W^{(0)}}^2)m_{W^{(k)}}^4+m_{u^{(0)}_\alpha} (-8m_{u^{(0)}_\alpha} (m_{u^{(0)}_\alpha}+m_{u^{(0)}_\beta})m_{W^{(0)}}^4-(m_{u^{(0)}_\alpha}-m_{u^{(0)}_\beta})^2 m_{u^{(0)}_\beta} (2m_{u^{(0)}_\alpha}+m_{u^{(0)}_\beta})
\nonumber \\ && \times
m_{W^{(0)}}^2+m_{(k)}^2((m_{u^{(0)}_\alpha}-m_{u^{(0)}_\beta})^2 m_{u^{(0)}_\beta}^2+8 m_{u^{(0)}_\alpha} (m_{u^{(0)}_\alpha}+m_{u^{(0)}_\beta})m_{W^{(0)}}^2))m_{W^{(k)}}^2+m_{u^{(0)}_\alpha}(m_{(k)}^2+m_{W^{(0)}}^2)
\nonumber \\ && \times
(m_{u^{(0)}_\alpha}^2 (m_{u^{(0)}_\alpha}-m_{u^{(0)}_\beta})^2 m_{u^{(0)}_\beta}^2+4 (m_{u^{(0)}_\alpha}+m_{u^{(0)}_\beta})(m_{u^{(0)}_\alpha}^3-m_{u^{(0)}_\beta}^2 m_{u^{(0)}_\alpha}+m_{u^{(0)}_\beta}^3)m_{W^{(0)}}^2)-2 m_{d^{(0)}_\gamma}^2m_{u^{(0)}_\alpha}^2
\nonumber \\ && \times
(m_{u^{(0)}_\alpha}+m_{u^{(0)}_\beta})(-2 m_{W^{(k)}}^4+4(m_{(k)}-m_{W^{(0)}}) (m_{(k)}+m_{W^{(0)}})m_{W^{(k)}}^2+(m_{u^{(0)}_\alpha}^2+m_{u^{(0)}_\beta}^2)(m_{(k)}^2+m_{W^{(0)}}^2))\big)\Big)
\nonumber \\ &&
+B_0(m^2_{u^{(0)}_\beta},m^2_{d^{(k)}_\gamma},m^2_{W^{(k)}})m_{u^{(0)}_\alpha} (m_{u^{(0)}_\alpha}-m_{u^{(0)}_\beta}) \Big(m_{u^{(0)}_\alpha} (m_{h^{(0)}}+m_{u^{(0)}_\alpha}-m_{u^{(0)}_\beta})(m_{h^{(0)}}-m_{u^{(0)}_\alpha}+m_{u^{(0)}_\beta})
\nonumber \\ && \times
(m_{(k)}^2+m_{W^{(0)}}^2)m_{W^{(k)}}^2 m_{d^{(k)}_\gamma}^4-\big(m_{u^{(0)}_\alpha}(-m^2_{h^{(0)}}+(m_{u^{(0)}_\alpha}-m_{u^{(0)}_\beta})^2)m_{W^{(k)}}^6+m_{u^{(0)}_\alpha}(m_{h^{(0)}}+m_{u^{(0)}_\alpha}-m_{u^{(0)}_\beta})
\nonumber \\ && \times
(m_{h^{(0)}}-m_{u^{(0)}_\alpha}+m_{u^{(0)}_\beta}) (3m_{(k)}^2-m_{W^{(0)}}^2)m_{W^{(k)}}^4+m_{u^{(0)}_\beta}(m_{(k)}^2+m_{W^{(0)}}^2)(m_{u^{(0)}_\alpha}^4-3 m_{u^{(0)}_\beta} m_{u^{(0)}_\alpha}^3
\nonumber \\ &&
+3 m_{u^{(0)}_\beta}^2 m_{u^{(0)}_\alpha}^2-m_{u^{(0)}_\beta}^3 m_{u^{(0)}_\alpha}-4m_{d^{(0)}_\gamma}^2 m_{u^{(0)}_\beta} m_{u^{(0)}_\alpha}+m_{h^{(0)}}^2 (m_{u^{(0)}_\beta}-m_{u^{(0)}_\alpha}) m_{u^{(0)}_\alpha}-4m_{d^{(0)}_\gamma}^2 m_{u^{(0)}_\beta}^2+4m_{u^{(0)}_\beta}
\nonumber \\ && \times
(m_{u^{(0)}_\alpha}+m_{u^{(0)}_\beta}) m_{W^{(0)}}^2)m_{W^{(k)}}^2+2 m_{h^{(0)}}^2 m_{u^{(0)}_\beta}^2 (m_{u^{(0)}_\alpha}+m_{u^{(0)}_\beta})(m_{(k)}^2+m_{W^{(0)}}^2)^2\big)m_{d^{(k)}_\gamma}^2+m_{h^{(0)}}^2
\nonumber \\ && \times
\big(-m_{u^{(0)}_\alpha}m_{W^{(k)}}^8+m_{u^{(0)}_\alpha}(2m_{(k)}^2+m_{u^{(0)}_\beta}^2-2m_{W^{(0)}}^2)m_{W^{(k)}}^6+m_{u^{(0)}_\beta}(m_{u^{(0)}_\alpha} (m_{u^{(0)}_\alpha}+2 m_{u^{(0)}_\beta})m_{W^{(0)}}^2-m_{(k)}^2
\nonumber \\ && \times
(m_{u^{(0)}_\alpha}^2+2 m_{u^{(0)}_\beta} m_{u^{(0)}_\alpha}+2 m_{u^{(0)}_\beta}^2))m_{W^{(k)}}^4+m_{u^{(0)}_\beta}(m_{(k)}^2+m_{W^{(0)}}^2)(4m_{u^{(0)}_\beta} (m_{u^{(0)}_\alpha}+m_{u^{(0)}_\beta})m_{(k)}^2+m_{u^{(0)}_\alpha}
\nonumber \\ && \times
(-m_{u^{(0)}_\alpha} m_{u^{(0)}_\beta}^2-4 (m_{u^{(0)}_\alpha}+m_{u^{(0)}_\beta}) m_{W^{(0)}}^2)) m_{W^{(k)}}^2+2m_{u^{(0)}_\alpha} m_{u^{(0)}_\beta}^3(m_{u^{(0)}_\alpha}+m_{u^{(0)}_\beta})(m_{(k)}^2+m_{W^{(0)}}^2)^2\big)
\nonumber \\ &&
+m_{W^{(k)}}^2 \big(m_{u^{(0)}_\alpha}(m_{u^{(0)}_\alpha}-m_{u^{(0)}_\beta})^2m_{W^{(k)}}^6-(m_{u^{(0)}_\alpha}(2m_{(k)}^2+m_{u^{(0)}_\beta}^2)(m_{u^{(0)}_\alpha}-m_{u^{(0)}_\beta})^2+2(-m_{u^{(0)}_\alpha}^3
\nonumber \\ &&
+2 m_{u^{(0)}_\beta} m_{u^{(0)}_\alpha}^2+m_{u^{(0)}_\beta}^2m_{u^{(0)}_\alpha}+2 m_{u^{(0)}_\beta}^3) m_{W^{(0)}}^2)m_{W^{(k)}}^4+m_{u^{(0)}_\beta} (-8m_{u^{(0)}_\beta} (m_{u^{(0)}_\alpha}+m_{u^{(0)}_\beta})m_{W^{(0)}}^4
\nonumber \\ &&
-m_{u^{(0)}_\alpha}(m_{u^{(0)}_\alpha}-m_{u^{(0)}_\beta})^2(m_{u^{(0)}_\alpha}+2 m_{u^{(0)}_\beta})m_{W^{(0)}}^2+m_{(k)}^2(m_{u^{(0)}_\alpha}^2 (m_{u^{(0)}_\alpha}-m_{u^{(0)}_\beta})^2+8 m_{u^{(0)}_\beta}(m_{u^{(0)}_\alpha}
\nonumber \\ &&
+m_{u^{(0)}_\beta})m_{W^{(0)}}^2))m_{W^{(k)}}^2+m_{u^{(0)}_\beta}(m_{(k)}^2+m_{W^{(0)}}^2)(m_{u^{(0)}_\alpha}^2 (m_{u^{(0)}_\alpha}-m_{u^{(0)}_\beta})^2 m_{u^{(0)}_\beta}^2+4 (m_{u^{(0)}_\alpha}+m_{u^{(0)}_\beta})(m_{u^{(0)}_\alpha}^3
\nonumber \\ &&
-m_{u^{(0)}_\beta} m_{u^{(0)}_\alpha}^2+m_{u^{(0)}_\beta}^3)m_{W^{(0)}}^2)-2 m_{d^{(0)}_\gamma}^2m_{u^{(0)}_\beta}^2 (m_{u^{(0)}_\alpha}+m_{u^{(0)}_\beta})(-2 m_{W^{(k)}}^4+4(m_{(k)}-m_{W^{(0)}}) (m_{(k)}
\nonumber \\ &&
+m_{W^{(0)}})m_{W^{(k)}}^2+(m_{u^{(0)}_\alpha}^2+m_{u^{(0)}_\beta}^2)(m_{(k)}^2+m_{W^{(0)}}^2)))\Big)
\bigg].
\end{eqnarray}
\end{widetext}

\section*{References}


\begin{thebibliography}{99}
%
\bibitem{Glashow} S. L. Glashow, {\it Partial-symmetries of weak interactions}, Nucl. Phys. {\bf 22}, 579 (1961).
%
\bibitem{Salam} A. Salam. {\it Weak and electromagnetic interactions}, Conf. Proc. {\bf C680519}, 367 (1968).
%
\bibitem{Weinberg} S. Weinberg, {\it A Model of Leptons}, Phys. Rev. Lett. {\bf 19}, 1264 (1967).
%
\bibitem{PDG} M. Tanabashi {\it et al.} (Particle Data Group), {\it Review of Particle Physics}, Phys. Rev. D {\bf 98}, 030001 (2018).
%
\bibitem{ATLASHiggs} G. Aad {\it et al} (The ATLAS Collaboration), {\it Observation of a new particle in the search for the Standard Model Higgs boson with the ATLAS detector at the LHC}, Phys. Lett. B {\bf 716}, 1 (2012).
%
\bibitem{CMSHiggs} S. Chatrchyan {\it et al} (The CMS Collaboration), {\it Observation of a new boson at mass of 125 GeV with the CMS experiment at the LHC}, Phys. Lett. B {\bf 716}, 30 (2012).
%
\bibitem{ATLASHb2013} G. Aad {\it et al} (The ATLAS Collboration), {\it Measurements of Higgs boson production and couplings in diboson final states with the ATLAS detector at the LHC}, Phys. Lett. B {\bf 726}, 88 (2013).
%
\bibitem{CMSHb2013} S. Chatrchyan {\it et al} (The CMS Collaboration), {\it Observation of a new boson with mass near 125 GeV in pp collisions at $\sqrt{s}=$7 and 8 TeV}, JHEP {\bf 1306}, 081 (2013).
%
\bibitem{EllYou} J. Ellis and T. You, {\it Updated global analysis of Higgs couplings}, JHEP {\bf 1306}, 103 (2013).
%
\bibitem{DjMo} A. Djouadi and G. Moreau, {\it The couplings of the Higgs boson and its CP properties from fits of the signal strengths and their ratios at the 7+8 TeV LHC}, Eur. Phys. J. C {\bf 73}, 2512 (2013).
%
\bibitem{ATLASHb2015} G. Aad {\it et al.} (The ATLAS Collaboration), {\it Study of the spin and parity of the Higgs boson in diboson decays with the ATLAS detector}, Eur. Phys. J. C {\bf 75}, 476 (2015).
%
\bibitem{ATLASandCMSHb2015} G. Aad {\it et al.} (The ATLAS and CMS Collaborations), {\it 	
Measurements of the Higgs boson production and decay rates and constraints on its couplings from a combined ATLAS and CMS analysis of the LHC pp collision data at $\sqrt{s}$=7 adn 8 TeV}, JHEP {\bf 1608}, 045 (2016).
%
\bibitem{ATLASHb2020} G. Aad {\it et al} (The ATLAS Collaboration), {\it Combined measurements of Higgs boson production and decay using up to 80 fb$^{-1}$ of proton-proton collision data at $\sqrt{s}$=13 TeVcollected with the ATLAS experiment}, Phys. Rev. D {\bf 101}, 012002 (2020).
%
\bibitem{Nordstrom} G. Nordstr\"om, {\it On the possibility of unifying the electromagnetic and gravitational fields}, Phys. Z. {\bf 15}, 504 (1914).
%
\bibitem{Kaluza} T. Kaluza, {\it Zum Unit\"atsproblem der Physik}, Sitzungsber. Preuss. Akad. Wiss. Berlin (Math. Phys.) {\bf 1921}, 966 (1921).
%
\bibitem{Klein} O. Klein, {\it Quantentheorie und f\"unfdimensionale Relativit\"atstheorie}, Z. Phys. {\bf 37} 895 (1926).
%
\bibitem{Veneziano} G. Veneziano, {\it Construction of a crossing-simmetric, Regge-behaved amplitude for linearly rising trajectories}, Nuovo Cim. A {\bf 57}, 190 (1968).
%
\bibitem{Nielsen}  H. B. Nielsen, {\it An Almost Physical Interpretation of the Dual N-Point Function}, Nordita Report, unpublished 1969.
%
\bibitem{KoNi1} Z. Koba and H. B. Nielsen, {\it Reaction amplitude for n-mesons a generalization of the Veneziano-Bardak\c{c}i-Ruegg-Virasoro model}, Nucl. Phys. B {\bf 10}, 633 (1969).
%
\bibitem{KoNi2} Z. Koba and H. B. Nielsen, {\it Manifestly crossing-invariant parametrization of n-meson amplitude}, Nucl. Phys. B {\bf 12}, 517 (1969).
%
\bibitem{Nambu}  Y. Nambu, {\it Quark Model and the Factorization of the Veneziano Amplitude}, in Symmetries and Quark Models, ed. R. Chand, p. 269, Gordon and Breach NY, 1970; Lectures at Copenhagen Symposium, 1970.
%
\bibitem{Susskind1} L. Susskind, {\it Structure of Hadrons Implied by Duality}, Phys. Rev. D {\bf 1}, 1182 (1970). 
%
\bibitem{Susskind2} L. Susskind, {\it Dual-symmetric theory of hadrons.-I}, Nuovo. Cim. A {\bf 69}, 457 (1970).
%
\bibitem{Lovelace} C. Lovelace, {\it Pomeron form factors and dual Regge cuts}, Phys. Lett B {\bf 34}, 500 (1971).
%
\bibitem{Ramond} P. Ramond, {\it Dual Theory for Free Fermions}, Phys. Rev. D {\bf 3}, 2415 (1971).
%
\bibitem{WeZu} J. Wess and B. Zumino, {\it Supergauge transformations in four dimensions}, Nucl. Phys. B {\bf 70}, 39 (1974).
%
\bibitem{SchSch} J. Scherk and J. H. Schwarz, {\it Dual models for non-hadrons}, Nucl. Phys. B {\bf 81}, 118 (1974).
%
\bibitem{Schwarz} J. H. Schwarz, {\it Physical states and pomeron poles in the dual pion model}, Nucl. Phys. B {\bf 46}, 61 (1972).
%
\bibitem{GS} M. B. Green and J. H. Schwarz, \textit{Anomaly cancellations in supersymmetric D =10 gauge theory and superstring theory}, Phys. Lett. B 149 (1984) 117.
%
\bibitem{GS1} M. B. Green and J. H. Schwarz, {\it Supersymmetrical string theories}, Phys. Lett. B {\bf 109}, 444 (1982).
%
\bibitem{GHMR1} D. J. Gross, J. A. Harvey, E. Martinec, and R. Rohm, {\it Heterotic String}, Phys. Rev. Lett. {\bf 54}, 502 (1985).
%
\bibitem{GHMR2} D. J. Gross, J. A. Harvey, E. Martinec, and R. Rohm, {\it Heterotic string theory (I). The free heterotic string}, Nucl. Phys. B {\bf 256}, 253 (1985).
%
\bibitem{GHMR3} D. J. Gross, J. A. Harvey, E. Martinec, and R. Rohm, {\it Heterotic string theory: (II). The interacting heterotic string}, Nucl. Phys. B {\bf 267}, 75 (1986).
%
\bibitem{Yau} S. -T. Yau, {\it On the Ricci curvature of a compact K\"ahler manifold and the complex Monge-Amp\`ere equation. I}, Commun. Pure Appl. Math. {\bf 31}, 339 (1978).
%
\bibitem{CHSW} P. Candelas, G. T. Horowitz, A. Strominger, and E. Witten, {\it Vacuum configurations for superstrings}, Nucl. Phys. B {\bf 258}, 46 (1985).
%
\bibitem{Witten} E. Witten, {\it String theory dynamics in various dimensions}, Nucl. Phys. B {\bf 443}, 85 (1995).
%
\bibitem{Polchinski} J. Polchinski, \textit{Dirichlet Branes and Ramond-Ramond Charges}, Phys. Rev. Lett. \textbf{75}, 4724 (1995).
%
\bibitem{HoWi1} P. Ho\v rava and E. Witten, {\it Heterotic and Type I string dynamics from eleven dimensions}, Nucl. Phys. B {\bf 460}, 506 (1996).
%
\bibitem{HoWi2} P. Ho\v rava and E. Witten, {\it Eleven-dimensional supergravity on a manifold with boundary}, Nucl. Phys. B {\bf 475}, 94 (1996).
%
\bibitem{Maldacena} J. Maldacena, {\it The Large-N Limit of Superconformal Field Theories and Supergravity}, Int. J.of Theor. Phys. {\bf 38}, 1113 (1999).
%
\bibitem{A} I. Antoniadis, Phys. Lett. B \textbf{246}, 377 (1990).
%
\bibitem{ADD}N. Arkani-Hamed, S. Dimopoulos, and G. R. Dvali, Phys.
Lett. B \textbf{429}, 263 (1998).
%
\bibitem{AADD}I. Antoniadis, N. Arkani-Hamed, S. Dimopoulos, and
G. R. Dvali, Phys. Lett. B \textbf{436}, 257 (1998).
%
\bibitem{RS1} L. Randall and R. Sundrum, {\it Large Mass Hierarchy from a Small Extra Dimension}, Phys. Rev. Lett. {\bf 83}, 3370 (1999).
%
\bibitem{RS2} L. Randall and R. Sundrum, {\it An Alternative to Compactification}, Phys. Rev. Lett. {\bf 83}, 4690 (1999).
%
\bibitem{ACD1} T. Appelquist, H. -C. Cheng, and B. Dobrescu, {\it Bounds on universal extra dimensions}, Phys. Rev. D {\bf 64}, 035002 (2001).
%
\bibitem{AppDo} T. Appelquist and B. A. Dobrescu, {\it Universal extra dimensions and the muon magnetic moment}, Phys. Lett. B {\bf 516}, 85 (2001).
%
\bibitem{RizzoUED} T. Rizzo, {\it Probes of universal extra dimensions at colliders}, Phys. Rev. D {\bf 64}, 095010 (2001).
%
\bibitem{ADPY} T. Appelquist, B. A. Dobrescu, E. Pont\'on, and H. -U. Yee, {\it Neutrinos vis-\`a-vis the six-dimensional standard model}, Phys. Rev. D {\bf 65}, 105019 (2002).
%
\bibitem{CMS} H. -C. Cheng, K. T. Matchev, and M. Schmaltz, {\it Bosonic supersymmetry? Getting fooled at the CERN LHC}, Phys. Rev. D {\bf 66}, 056006 (2002).
%
\bibitem{SeTa1} G. Servant and T. M. P. Tait, {\it Elastic scattering and direct detection of Kaluza-Klein dark matter}, New J. Phys. {\bf 4}, 99 (2002).
%
\bibitem{SeTa2} G. Servant and T. M. P. Tait, {\it Is the lightest Kaluza-Klein particle a viable dark matter candidate?}, Nucl. Phys. B {\bf 650}, 391 (2003).
%
\bibitem{AppYee} T. Appelquist and H. -U. Yee, {\it Universal extra dimensions and the Higgs boson mass}, Phys. Rev. D {\bf 67}, 055002 (2003).
%
\bibitem{BSW} A. J. Buras, M. Spranger, and A. Weiler, {\it The impact of universal extra dimensions on the unitarity triangle and rare K and B decays}, Nucl. Phys. B {\bf 660}, 225 (2003).
%
\bibitem{BPSW} A. J. Buras, A. Poschenrieder, M. Spranger, and A. Weler, {\it The impact of universal extra dimensions on $B\to\chi_s\gamma$, $B\to\chi_sgluon$, $B\to\chi_s\mu^+\mu^-$, $K_L\to\pi^0e^+e^-$ and $\varepsilon'/\varepsilon$}, Nucl. Phys. B {\bf 678}, 455 (2004).
%
\bibitem{HoPr} D. Hooper and S. Profumo, {\it Dark matter and collider phenomenology of universal extra dimensions}, Phys. Rept. {\bf 453}, 29 (2007).
%
\bibitem{NT} H. Novales-S\' anchez and J. J. Toscano, {\it Gauge invariance and quantization of Yang-Mills theories in extra dimensions}, Phys. Rev. D \textbf{82}, 116012 (2010).
%
\bibitem{CGNT} A. Cordero-Cid, M. G\'omez-Bock, H. Novales-S\'anchez, and J. J. Toscano, {\it The Standard Model with one universal extra dimension}, Pramana {\bf 80}, 369 (2013).
%
\bibitem{Servant} G. Servant, {\it Status report on universal extra dimensions after LHC8}, Mod. Phys. Lett A {\bf 30}, 1540011 (2015). 
%
\bibitem{DFK} N. Deutschmann, T. Flacke, and J. S. Kim, {\it Current LHC constraints on minimal universal extra dimensions}, Phys. Lett B {\bf 771}, 515 (2017).
%
\bibitem{DDG} K. R. Dienes, E. Dudas, and T. Gherghetta, {\it Grand unification at intermediate mass scales through extra dimensions}, Nucl. Phys. B {\bf 537}, 47 (1999).
%
\bibitem{MPR} A. M\"uck, A. Pilaftsis, and R. R\"uckl, {\it Minimal higher-dimensional extensions of the standard model and electroweak observables}, Phys. Rev. D 65, 085037 (2002).
%
\bibitem{RizzoED} T. G. Rizzo, {\it introduction to Extra Dimensions}, AIP Conf. Proc. {\bf 1256}, 27 (2010).
%
\bibitem{BKP} G. B\'elanger, M. Kakizaki, and A. Pukhov, {\it Dark matter in UED: the role of the second KK level}, JCAP {\bf 2011}, 009 (2011).
%
\bibitem{FKKMP} T. Flacke, D. W. Kang, K. Kong, G. Mohlabeng, and S. C. Park, {\it Electroweak Kaluza-Klein dark matter}, JHEP {\bf 04}, 041 (2017).
%
\bibitem{LMNT2} M. A. L\' opez-Osorio, E. Mart\'\i nez-Pascual, H. Novales-S\' anchez, J. J. Toscano, {\it Yang-Mills theories with an arbitrary number of compactified extra dimensions} Phys. Rev. D \textbf{89}, 116015 (2014).
%
\bibitem{EnBr}F. Englert and R. Brout, \textit{Broken Symmetry and the Mass of Gauge Vector Mesons }, Phys. Rev. Lett. {\bf 13} 321 (1964).
%
\bibitem{PWHiggs1} P. W. Higgs, \textit{Broken symmetries, massless particles and gauge fields }, Phys. Lett. {\bf 12} 132 (1964).
%
\bibitem{PWHiggs2} P. W. Higgs, \textit{Broken Symmetries and the Masses of Gauge Bosons}, Phys. Rev. Lett. {\bf 13} 508 (1964).
%
\bibitem{MNT} J. Monta\~no, H. Novales-S\'anchez, and J. J. Toscano, {\it Effects of universal extra dimensions on top-quark electromagnetic interactions}, J. Phys. G {\bf 47}, 015002 (2019).
%
\bibitem{Petriello} F. Petriello, {\it Kaluza-Klein Effects on Higgs Physics in Universal Extra Dimensions}, JHEP {\bf 0205}, 003 (2002). 
%
\bibitem{NoTo1} H. Novales-S\'anchez and J. J. Toscano, {\it About gauge invariance in compactified extra dimensions}, Phys. Rev. D {\bf 84}, 057901 (2011).
%
\bibitem{BBBKP} G. B\'elanger, A. Belyaev, M. Brown, M. Kakizaki, and A Pukhov, {\it Testing minimal universal extra dimensions using Higgs boson searchas at the LHC}, Phys. Rev. D {\bf 87}, 016008 (2013).
%
\bibitem{DKandSR} U. K. Dey and T. S. Ray, {\it Constraining minimal and nonminimal universal extra dimension models with Higgs couplings}, Phys. Rev. D 88, 056016 (2013).
%
\bibitem{KDandJ} U. K. Dey and T. Jha, {\it Rare top decays in minimal and nonminimal universal extra dimension models}, Phys. Rev. D {\bf 94}, 056011 (2016).
%
\bibitem{PassVe} G. Passatino and M. Veltman, {\it One-loop corrections for $e^+e^-$ annihilation into $\mu^+\mu^-$ in the Weinberg model}, Nucl. Phys. B {\bf 160}, 151 (1979).
%
\bibitem{HooVe} G. 't Hooft and M. Veltman, {\it Scalar one-loop integrals}, Nucl. Phys. B {\bf 153}, 365 (1979).
%
\bibitem{LMNT1} M. A. L\' opez-Osorio, E. Mart\'\i nez-Pascual, H. Novales-S\' anchez, J. J. Toscano, {\it Hidden symmetries induced by a canonical transformation and gauge structure of compactified Yang-Mills theories}, Phys. Rev. D \textbf{88}, 036015 (2013).
%
\bibitem{CDH} H. -C. Cheng, B. A. Dobrescu, and C. T. Hill, {\it Electroweak symmetry breaking and extra dimensions}, Nucl.Phys. B {\bf 589}, 249 (2000).
%
\bibitem{PapaSan} J. Papavassiliou and A. Santamaria, {\it Chiral fermions and gauge fixing in five-dimensional theories}, Phys. Rev. D {\bf 63}, 125014 (2001).
%
\bibitem{HosoConf} Y. Hosotani, {\it New Dimensions from Gauge-Higgs Unification}, PoS {\bf CORFU2016}, 026 (2017).
%
\bibitem{GiKi} C. Giunti and C.W. Kim, {\it Fundamentals of Neutrino Physics and Astrophysics} (Oxford University Press, New York, 2007).
%
\bibitem{CheLi} T.-P. Cheng and L.-F. Li, {\it Gauge theory of elementary particle physics} (Oxford University Press, Oxford,1988).
%
\bibitem{PeSch} M. E. Peskin and D. V. Schroeder, {\it An Introduction to Quantum Field Theory} (Perseus, Reading, 1995).
%
\bibitem{HenTe} M. Henneaux and C. Teitelboim, {\it Quantization of Gauge Systems} (Princeton University Press, Princeton, 1992)
%
\bibitem{MRS} N. M. Monyonko, J. H. Reid, and A. Sen, {\it Some properties of green's functions in the non-linear $R_\xi$ gauge}, Phys. Lett. B {\bf 136}, 265 (1984).
%
\bibitem{MoRe} N. M. Monyonko and J. H. Reid, {\it One-loop vacuum polarization in the nonlinear $R_\xi$ gauge}, Phys. Rev. D {\bf 32}, 962 (1985).
%
\bibitem{RoBa} J. C. Romo and A. Barroso, {\it Renormalization of the electroweak theory in the nonlinear gauge}, Phys. Rev. D {\bf 35}, 2836 (1987).
%
\bibitem{MeTo} J. G. M\'endez and J. J. Toscano, {\it A nonlinear $R_\xi$ gauge for the elctroweak theory}, Rev. Mex. Fis. {\bf 50}, 346 (2004).
%
\bibitem{BRS1} C. Becchi, A. Rouet, and R. Stora, {\it Renormalization of the abelian Higgs-Kibble model}, Commun. Math. Phys. {\bf 42}, 127 (1975).
%
\bibitem{BRS2} C. Becchi, A. Rouet, and R. Stora, {\it Renormalization of gauge theories}, Ann. Phys. {\bf 98},
287 (1976).
%
\bibitem{Tyutin} I. V. Tyutin and P. N. Lebedev, Physical Institute of the USSR Academy of Science Report No. 39, 1975 (to be published).
%
\bibitem{GPS} J. Gomis, J. Paris, and S. Samuel, {\it Antibracket, antifields and gauge-theory quantization}, Phys. Rep. {\bf 259}, 1 (1995).
%
\bibitem{BaVi1} I. A. Batalin and G. A. Vilkovisky, {\it Gauge algebra and quantization}, Phys. Lett. B {\bf 102}, 27 (1981).
%
\bibitem{BaVi2} I. A. Batalin and G. A. Vilkovisky, {\it Feynman rules for reducible gauge theories}, Phys. Lett. B {\bf 120}, 166 (1983).
%
\bibitem{BaVi3} I. A. Batalin and G. A. Vilkovisky, {\it Quantization of gauge theories with linearly dependent generators}, Phys. Rev. D {\bf 28}, 2567 (1983); {\bf 30}, 508(E) (1984).
%
\bibitem{BaVi4} I. A. Batalin and G. A. Vilkovisky, {\it Closure of the gauge algebra, generalized lie equations and Feynman rules}, Nucl. Phys. B {\bf 234}, 106 (1984).
%
\bibitem{BaVi5} I. A. Batalin and G. A. Vilkovisky, {\it Existence theorem for gauge algebra}, J. Math. Phys.
{\bf 26}, 172 (1985).
%
\bibitem{FLS} K. Fujikawa, B. W. Lee, and A. I. Sanda, {\it Generalized Renormalizable Gauge Formulation of Spontaneously Broken Gauge Theories}, Phys. Rev. D {\bf 6}, 2923 (9172).
%
\bibitem{NT1} H. Novales-S\'anchez and J. J. Toscano, {\it Integration of Kaluza-Klein modes in Yang-Mills theories}, Phys. Rev. D {\bf 84}, 076010 (2011).
%
\bibitem{GNT} I. Garc\'ia-Jim\'enez, H. Novales-S\'anchez, and J. J. Toscano, {\it Distinctive ultraviolet structure of extra-dimensional Yang-Mills theories by integration of heavy Kaluza-Klein modes}, Phys. Rev. D {\bf 93}, 096007 (2016).
%
\bibitem{Boudjema} F. Boudjema, {\it The scattering of light by light in the non-linear gauge}, Phys. Lett. B {\bf 187}, 362 (1987).
%
\bibitem{HPTT} J. M. Hern\'andez, M. A. P\'erez, G. Tavares-Velasco, and J. J. Toscano, {\it Decay $Z\to\bar{\nu}\nu\gamma$ in the standard model}, Phys. Rev. D {\bf 60}, 013004 (1999).
%
\bibitem{HPTT1} J. Hern\'andez-S\'anchez, M. A. P\'erez, G. Tavares-Velasco, and J. J. Toscano, {\it Decay $H^+\to W^+\gamma$ in a nonlinear $R_\xi$ gauge}, Phys. Rev. D {\bf 69}, 095008 (2004).
%
\bibitem{MTTR} J. Monta\~no, G. Tavares-Velasco, J. J. Toscano, and F. Ram\'irez-Zavaleta, {\it $SU_L$(3)$\times U_X$(1)-invariant description of the bilepton contribution to the WWV vertex in the minimal 331 model}, Phys. Rev. D {\bf 72}, 055023 (2005).
%
\bibitem{RTT} F. Ram\'irez-Zavaleta, G. Tavares-Velasco, and J. J. Toscano, {\it Bilepton effects on the $WWV^*$ vertex in the 331 model with right-handed neutrinos via a $SU_L$(2)$\times U_Y$(1) covariant quantization scheme}, Phys. Rev. D {\bf 75}, 075008 (2007).
%
\bibitem{HoTo} C. G. Honorato and J. J. Toscano, {\it $U_e$(1)-covariant $R_\xi$ gauge for the two-Higgs doublet model}, Pramana {\bf 73}, 1023 (2009).
%
\bibitem{HHPT} J. Hern\'andez-S\'anchez, C. G. Honorato, M. A. P\'erez, and J. J. Toscano, {\it $\gamma\gamma\to\phi_i\phi_j$ processes in the type-III two-Higgs-doublet model}, Phys. Rev. D {\bf 85}, 015020 (2012).
%
\bibitem{Langacker} P. Langacker, {\it The Standard Model and Beyond} (Taylor \& Francis Group, Boca Raton, 2010).
%
\bibitem{NCabibbo} N. Cabibbo, {\it Unitary Symmetry and Leptonic Decays}, Phys. Rev. Lett. {\bf 10}, 531 (1963).
%
\bibitem{KoMa} M. Kobayashi and T. Maskawa, {\it CP-Violation in the Renormalizable Theory of Weak Interaction}, Prog. Theor. Phys. {\bf 49}, 652 (1973).
%
\bibitem{Wolfenstein} L. Wolfenstein, {\it Parametrization of the Kobayashi-Maskawa Matrix}, Phys. Rev. Lett. {\bf 51}, 1945 (1983).
%
\bibitem{ChKe} L.-L. Chau and W.-Y. Keung, {\it Comments on the Parametrization of the Kobayashi-Maskawa Matrix}, Phys. Rev. Lett. 53, 1802 (1984).
%
\bibitem{Sakharov} A. D. Sakharov, {\it Violation of CP invariance, C asymmetry, and baryon asymmetry of the universe}, Pis’ma Zh. Eksp. Teor. Fiz. {\bf 5}, 32 (1967) [JETP Lett. {\bf 5}, 24 (1967)] [Violation ofCPin variance,Casymmetry, and baryon asymmetry of the universe, Sov. Phys. Usp. {\bf 34}, 392 (1991)] [Usp. Fiz. Nauk {\bf 161}, 61 (1991)].
%
\bibitem{BGLMT} L. G. Benitez-Guzm\'an, I. Garc\'ia-Jim\'enez, M. A. L\'opez-Osorio, E. Mart\'inez-Pascual, and J. J. Toscano, {\it Revisiting the flavor changing neutral current Higgs decays $H\to q_iq_j$ in the Standard Model}, Journal of Physics G {\bf 42}, 085002 (2015).
%
\bibitem{GAMRT} G. Gonz\'alez-Estrada, J. I. Aranda, J. Monta\~no, F- I. Ram\'irez-Zavaleta, and E. S. Tututi, {\it The rare $H\to q_iq_j$ decays revisited}, PoS {\bf LHCP2019}, 066 (2019).
%
\bibitem{BoGi} C. G. Bollini and J. J. Giambiagi, {\it Dimensional renormali- zation: The number of dimensions as a regularizing param- eter}, Nuovo Cim. B {\bf 12}, 20 (1972).
%
\bibitem{MBD} R. Mertig, M. B\"ohm, and A. Denner, {\it Feyn Calc - Computer-algebraic calculation of Feynman amplitudes}, Comput. Phys. Commun. {\bf 64}, 345 (1991).
%
\bibitem{GIMmech} S. L. Glashow, J. Iliopoulos, and L. Maiani, {\it Weak Interactions with Lepton-Hadron Symmetry}, Phys. Rev. D {\bf 2}, 1285 (1970).
%
\bibitem{MNNST} E. Mart\'inez-Pascual, G. N\'apoles-Ca\~nedo, H. Novales-S\'anchez, A. Sierra-Mart\'inez, and J. J. Toscano, {\it Implications of extra dimensions on the effective charge and beta function in quantum electrodynamics}, Phys. Rev. D {\bf 101}, 035034 (2020).
%
\bibitem{HNT} M. Huerta-Leal, H. Novales-S\'anchez, and J. J Toscano, {\it A gauge-invariant approach to asymptotic freedom in Yang-Mills theories with universal extra dimensions}, submitted to {\it Phys. Rev. D}, ePrint: arXiv:1704.07339 (2020).
%
\bibitem{LMNTo} M. A. L\'opez-Osorio, E. Mart\'inez-Pascual, G. I. N\'apoles-Ca\~nedo, and J. J. Toscano, {\it One-loop order effects from one universal extra dimension on $\lambda\phi^4$ theory}, submitted to {\it J. Phys. G} (2020).
%
\bibitem{GMNNTT} I. Garc\'ia-Jim\'enez, J. Monta\~no, G. N\'apoles-Ca\~nedo, H. Novales-S\'anchez, J. J. Toscano, and E. S. Tututi, {\it Diphoton Higgs signal strength in universal extra dimensions}, submitted to {\it J. Phys. G}, ePrint: arXiv:1705.02637 (2020).
%
\bibitem{Epstein} P. Epstein, {\it Zur Theorie allgemeiner Zetafunctionen}, Math. Annalen {\bf 56}, 615 (1903).
%
\bibitem{Weinbergbook} S. Weinberg, {\it The Quantum Theory of Fields} (Cambridge University Press, New York, 1995).
%
\bibitem{Wudka} J. Wudka, {\it Electroweak effective lagrangians}, Int. J. Mod. Phys. A {\bf 9}, 2301 (1994).
%
\bibitem{BuWy} W. Buchm\"uller and D. Wyler, {\it Effective lagrangian analysis of new interactions and flavour conservation}, Nucl. Phys. B {\bf 268}, 621 (1986).
%
\bibitem{LLR} C. N. Leung, S. T. Love, and S. Rao, {\it Low-energy manifestations of a new interactions scale: Operator analysis}, Z. Phys. C - Particles and Fields {\bf 31}, 433–437 (1986).
%
\bibitem{AppCar} T. Appelquist and J. Carazzone, {\it Infrared singularities and massive fields}, Phys. Rev. D 11, 2856 (1975).
%
\bibitem{HaPe} T. Hahn and M. P\'erez-Victoria, {\it Automated one-loop calculation in four and D dimensions}, Comput. Phys. Commun. {\bf 118}, 153 (1999).
%
\bibitem{OldVer} G. J. Oldenborgh and J. A. M. Vermaseren, {\it New algorithms for one-loop integrals}, Z. Phys. C {\bf 46}, 425 (1990).
%
\bibitem{BEI} G. Blankenburg, J. Ellis, and G. Isidori, {\it Flavour-changing decays of a 125 GeV Higgs-like particle}, Phys. Lett. B {\bf 712}, 386 (2012).
%
\bibitem{HKZ} R. Harnik, J. Kopp, and J. Zupan, {\it Flavor violating Higgs decays}, JHEP {\bf 1303}, 026 (2013).
%
\bibitem{EHS} G. Eilam, J. L. Hewett, and A. Soni, {\it Rare decays of the top quark in the standard and two-Higgs-doublet models}, Phys. Rev. D {\bf 44}, 1473 (1991).
%
\bibitem{BDDM} J. Beuria, A. Datta, D. Debnath, and K. T. Matchev, {\it LHC collider phenomenology of minimal universal extra dimensions}, Comput. Phys. Commun. {\bf 226}, 187 (2018).
%
\bibitem{HaWe} U. Haisch and A. Weiler, {\it Bound on minimal universal extra dimensions from $\bar{B}\to X_s\gamma$}, Phys. Rev. D {\bf 76}, 034014 (2007).
%
\bibitem{DGKN} G. Dvali, G. Gabadadze, M. Kolanovi\'c, and F. Nitti, {\it Power of brane-induced gravity}, Phys. Rev. D {\bf 64}, 084004 (2001).
%
\bibitem{CTW} M. Carena, T. M. P. Tait, and C. E. M. Wagner, {\it Branes and orbifolds are opaque}, Acta Phys. Polon. B {\bf 33}, 2355 (2002).
%
\bibitem{APS} F. del Aguila, M. P\'erez-Victoria, and J. Santiago, {\it Effective description of brane terms in extra dimensions}, JHEP {\bf 0610}, 056 (2006).
%
\end{thebibliography}

\end{document}